\documentclass[prd,preprint,tightenlines,showpacs,preprintnumbers,nofootinbib,eqsecnum,superscriptaddress]{revtex4-1}

 \usepackage[dvips,final]{graphicx}
  \usepackage{amssymb}
   \usepackage{amsmath}
    \usepackage{amsfonts}
     \usepackage{epsfig}
      \usepackage{bm}

\usepackage[section]{placeins}

\usepackage{sidecap}
\usepackage{multirow}
\usepackage{booktabs}
\usepackage{array}
\usepackage{tabularx}
\usepackage{xcolor}
\usepackage{pstricks}
\def\frac#1#2{{\begingroup #1\endgroup\over #2}}

\begin{document}

\title{Production of $e^{+} e^{-}$ in proton-lead collision: photon-photon fusion}

\author{Barbara Linek}
\email{barbarali@dokt.ur.edu.pl}
\affiliation{College of Natural Sciences, Institute of Physics, University of Rzeszów,
ul. Pigonia 1, PL-35-959 Rzeszów, Poland}
\author{Marta Łuszczak}
\email{mluszczak@ur.edu.pl}
\affiliation{College of Natural Sciences, Institute of Physics, University of Rzeszów,
ul. Pigonia 1, PL-35-959 Rzeszów, Poland}
\author{Wolfgang Sch\"afer}
\email{Wolfgang.Schafer@ifj.edu.pl} 
\affiliation{Institute of Nuclear
Physics, Polish Academy of Sciences, ul. Radzikowskiego 152, PL-31-342 
Krak{\'o}w, Poland}
\author{Antoni Szczurek}
\email{antoni.szczurek@ifj.edu.pl}
\affiliation{Institute of Nuclear Physics, Polish Academy of Sciences, 
ul. Radzikowskiego 152, PL-31-342 Krak{\'o}w, Poland}

\vspace{0.9cm}

\begin{abstract}
We analyze the photon-initiated processes for production of $ e ^ + e ^- $ pairs in proton-nucleus collisions at LHC energy, taking into account both elastic processes and proton dissociations in the low-mass region (LMR) and intermediate-mass region (IMR) as defined by the ALICE collaboration. The calculations are performed within the $k_{\rm T}$-factorisation approach, including transverse momenta of intermediate photons. We discuss several differential distributions in invariant mass of both the leptons $M_{ll}$, pair rapidity $Y_{ll}$ and transverse momenta of the lepton pair $p_{t,ll}$. In addition, we present the two-dimensional distributions in $\log_{10} x_{Bj}$ and $\log_{10} Q^{2}$ and ($\log_{10} W,\log_{10} Q^{2}$) the arguments of the deep-inelastic structure functions. All presented results were obtained with modern parametrizations of proton structure functions. Limiting to small invariant masses of dielectrons one tests structure functions in the nonperturbative region of small $Q^2$ and/or small $W$. We quantify difference for different parametrizations from the literature.
\end{abstract}

\maketitle

\section{Introduction}

There are many mechanisms of dilepton production in $p p$, $p A$ and $A A$ collisions. In $p p$ collisions these are Dalitz decays at low dilepton masses and semileptonic decays of mesons or Drell-Yan processes. There the gamma-gamma processes exist but the contribution is rather small. However, it can be measured by imposing rapidity gaps, see for example \cite{Forthomme:2018sxa}. Till recently in nucleus-nucleus collision the problem was separated into real hadronic collisions ($b < R_1 + R_2$), where the mechanism are similar as in proton-proton collisions and ultraperipheral collisions ($b > R_1 + R_2$), where the dominant mechanism is photon-photon fusion. It was, however, shown, (see e.g. \cite{Klusek-Gawenda:2018zfz}) that the photon-photon processes survive also in the semi-central collisions and actually dominate at very small transverse momenta of the dilepton pair. In $p A$ collisions the issue was not carefully analysed. The only exception is \cite{Dyndal:2019ylt}. The authors of the paper made a feasibility study for the ATLAS experimental apparatus.  
The recent ALICE measurements~\cite{ALICE:2020mfy} on dilepton production in proton-lead collisions are the motivation for carrying out the present research.

Here we wish to thoroughly investigate the contribution of photon-initiated processes to the production of dileptons in proton-nucleus collisions, in order to determine the parameters enabling future measurements. Due to the fact that the nucleus in discussed collisions is only a source of "elastic" photons, there are only two types of photon-initiated dilepton production in proton-nucleus collisions for energy 5.02 TeV, which are called doubly-elastic and single dissociation.

Dilepton production in $pA$-collisions with a rapidity gap between the nucleus and a high-$p_T$ lepton has been suggested as a probe of the photon partonic content of the proton \cite{Dyndal:2019ylt}.

Photons as partons of the proton are attracting much attention recently \cite{Luszczak:2015aoa,Manohar:2016nzj,xFitterDevelopersTeam:2017fxf,Harland-Lang:2019pla,Luszczak:2018ntp}, as they can play an important role in a number of electroweak processes. They are especially important in event topologies with rapidity gaps as for example in  \cite{Forthomme:2018sxa,Luszczak:2018dfi}, but can also have a significant contribution to precise determination of inclusive observables, see e.g. \cite{Denner:2019vbn}.

\section{Formalism}

\begin{figure}
  \centering
  \includegraphics[width=.325\textwidth]{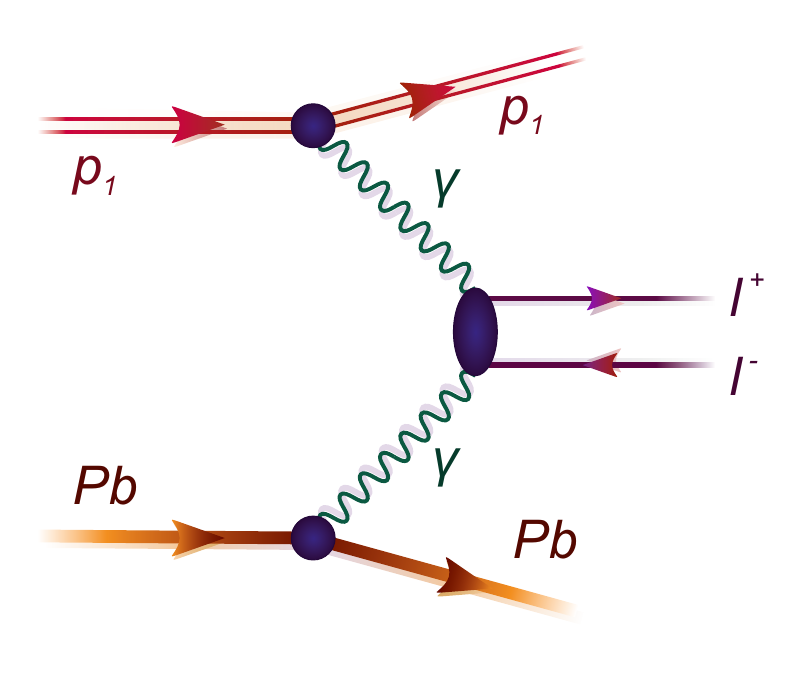}
  \includegraphics[width=.325\textwidth]{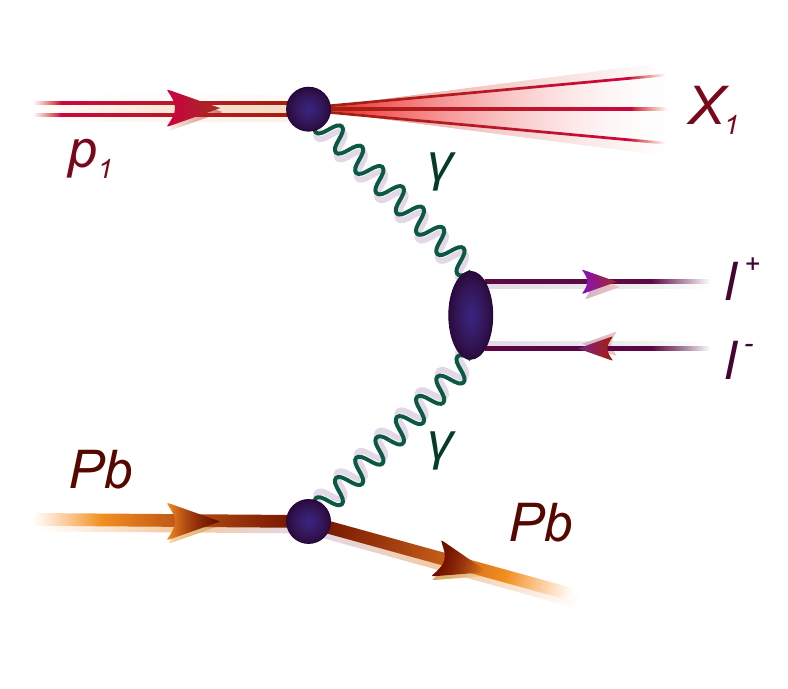}
  \caption{Classes of processes discussed in the present
paper. From left to right: elastic-elastic, inelastic-elastic (or equivalently, elastic-inelastic).}
\label{fig:diagrams}
\end{figure}

Fig.\ref{fig:diagrams} shows schematically diagrams of processes included in our present analysis. In the present paper we concentrate on general characteristics and study of differential distribution to select a proper observable for future experimental studies.

\subsection{Fluxes of elastic photons}

To obtain distributions of elastic photons from a proton, it is necessary to express the equivalent photon flux using electric $G_{E}(Q^2)$ and magnetic $G_{M}(Q^2)$ form factors, what is expressed as
\begin{eqnarray}
Q^2 {d \gamma^{p}_{el}(x,Q^2) \over dQ^2} = \frac{\alpha_{\rm{em}}}{\pi}
 \Big[ \Big( 1- {x \over 2} \Big)^2 \frac{4 m_p^2 G_E^2(Q^2) + Q^2 G_M^2(Q^2)}{4m_p^2 + Q^2} + {x^2 \over 4} G_M^2(Q^2) \Big]~,
\label{proton_el_flux}
\end{eqnarray}
where $x$ is the fraction of the proton's momentum carried by the photon and $m_{p}$ is the mass of the proton.

In order to express the elastic photon flux for the nucleus $(\gamma_{el}^{Pb})$ in accordance with Ref.~\cite{Budnev:1974de} we replaced 
\begin{eqnarray}
 {4 m_p^2 G_E^2(Q^2) + Q^2 G_M^2(Q^2) \over 4m_p^2 + Q^2} \longrightarrow Z^2 F_{\rm em}^2(Q^2)~,
 \end{eqnarray}
where $Z$ is the charge of the nucleus and $F_{\rm em}(Q^2)$ is its charge form factor.

In the case of the $^{208}$Pb nucleus, we used the form factor parametrization used in the STARlight MC generator~\cite{Klein:2016yzr}:
\begin{eqnarray}
 F_{\rm em}(Q^2) = {3 \over (QR_A)^3}\Big[ \sin(QR_A) - QR_A \cos(QR_A) \Big] { 1 \over 1 + a^2 Q^2}~,
 \label{pb_el_flux}
\end{eqnarray}
where $Q = \sqrt{Q^2}$, $R_A = 1.1 A^{1/3}$ fm and $a = 0.7$ fm, and $A=208$, $Z=82$.

Integrating the elastic photon PDFs of the proton and the lead nucleus over $Q^2$ we have
\begin{equation}
\gamma^{(p,Pb)}_{el}(x)  = \int_{Q^2_{\rm min}} dQ^2  {d \gamma^{(p,Pb)}_{el}(x, Q^2)
\over d Q^2}\, , \qquad Q^2_{\rm min} = {x^2 m_p^2 \over 1-x} \, . 
 \label{integral_flux}
\end{equation}

\subsection{High-energy factorization}
In this paper, the $k_{T}$-factorization approach, called also high-energy factorization, is used, in which the vertices $\gamma^*p \rightarrow X$ can be parameterized in terms of the proton structure function. Photons from inelastic cases are characterized in this approach by having transverse momenta and non-zero virtuality, and by using unintegrated photon distributions. These fluxes, in the DIS limit, can be calculated from the equation (see e.g.~\cite{daSilveira:2014jla,Luszczak:2015aoa}):
\begin{eqnarray}
Q^2 {d \gamma^p_{inel}(x,Q^2) \over dQ^2} = {1\over x} \, 
\int_{M^2_{\rm thr}} dM_X^2 {\cal{F}}^{\mathrm{in}}_{\gamma^* \leftarrow p} (x,\vec{q}_T^2,M^2_X) \, ,
\end{eqnarray}
using functions $ {\cal{F}}^{\mathrm{in}}_{\gamma^* \leftarrow p}$ from \cite{Budnev:1974de, Luszczak:2018ntp}:
\begin{eqnarray}
{\cal{F}}^{\mathrm{in}}_{\gamma^* \leftarrow p} (x,\vec{q}_T^2, M_X) &=& {\alpha_{\rm em} \over \pi} 
\Big\{(1-x) \Big( {\vec{q}_T^2 \over \vec{q}_T^2 + x (M_X^2 - m_p^2) + x^2 m_p^2  }\Big)^2  
{F_2(x_{\rm Bj},Q^2) \over Q^2 + M_X^2 - m_p^2}  \nonumber \\
&+& {x^2 \over 4 x^2_{\rm Bj}}  
{\vec{q}_T^2 \over \vec{q}_T^2 + x (M_X^2 - m_p^2) + x^2 m_p^2  }
{2 x_{\rm Bj} F_1(x_{\rm Bj},Q^2) \over Q^2 + M_X^2 - m_p^2} \Big\} \, ,
\label{eq:flux_in}
\end{eqnarray}
 where
 \begin{eqnarray}
Q^2 =  {\vec{q}_T^2 + x (M_X^2 - m_p^2) + x^2 m_p^2 \over (1-x)} \, ,
\label{eq:q2}
\end{eqnarray}
and
\begin{eqnarray}
x_{\rm Bj} = {Q^2 \over Q^2+M^2_X -m_p^2}.
\end{eqnarray}
In practice, we use the functions $F_{L}(x_{\rm Bj},Q^2)$ and $F_{2}(x_{\rm Bj},Q^2)$ instead of $F_{1}(x_{\rm Bj},Q^2)$ and $F_{2}(x_{\rm Bj},Q^2)$. The $F_{L}(x_{\rm Bj},Q^2)$ function, which is the proton's longitudinal structure function, can be expressed by the functions $F_{1}(x_{\rm Bj},Q^2)$ and $F_{2}(x_{\rm Bj},Q^2)$ as:
\begin{eqnarray}
F_L(x_{\rm Bj},Q^2) = \Big( 1 + {4 x_{\rm Bj}^2 m_p^2 \over Q^2} \Big) F_2(x_{\rm Bj},Q^2) - 2 x_{\rm Bj} F_1(x_{\rm Bj},Q^2).
\end{eqnarray}
Therefore, in the $k_{T}$-factorization approach, the cross-section for the $p+Pb\rightarrow Pb+l^+l^-+X$ processes is (taking into account unintegrated photon flux):
\begin{equation}
\sigma = S^2 \int dx_p dx_{\rm Pb} {d^2\vec{q}_T \over \pi} \Big[ {d \gamma^{p}_{el}(x_p, Q^2) \over dQ^2} + { d \gamma^{p}_{inel}(x_p,Q^2) \over dQ^2} \Big]
 \gamma^{\rm Pb}_{el}(x_{\rm Pb})
\sigma_{\gamma^{*}  \gamma \rightarrow \ell^+ \ell^-}(x_p, x_{\rm Pb}, \vec{q_T}) \,,
\label{kt_factorization_formula}
\end{equation}
where $\sigma_{\gamma^{*} \gamma \rightarrow \ell^+ \ell^-}$ is the off-shell elementary cross-section (for details see Refs.~\cite{daSilveira:2014jla,Catani:1990eg}) and  for  $x_p \ll 1$ we can assume that $Q^2 \approx \vec{q}_T^2$ (see Eq.(~\ref{eq:q2})).

Here we also put a gap-survival factor $S^2 \leq 1$ in front. In fact the gap survival probability is expected to depend on the kinematics of the process. It should be applied when asking for a rapidity gap.
The modelling of the latter goes beyond the scope of this work. Furthermore, we concentrate on the contribution to inclusive observables, where $S^2 = 1$.  

Importantly, despite the fact that the fluxes do not depend on the $\vec{q}_T$ direction, for collinear case, the averaging over $\vec{q}_T$ directions in the off-shell cross-sections replaces the average of photon polarization.

\subsection{Structure function parametrizations}

We expect, that in the kinematical region of interest in this work, the main contribution will come 
from the structure functions probed in the nonperturbative region, where their $Q^2$ and $x_{\rm Bj}$ dependence cannot be calculated by perturbative QCD. To control the inevitable model dependence, we use a variety of structure function parametrizations. Three of them, the ALLM~\cite{Abramowicz:1991xz,Abramowicz:1997ms}, FFJLM (Fiore et al.)~\cite{Fiore2002ExplicitMR} and LUX-like~\cite{Manohar:2017eqh} were already used in our previous publications and are described in more detail in Ref.~\cite{Luszczak:2018ntp}.
A new addition in this work is a parametrization by Kulagin and Barinov \cite{Kulagin:2021mee}. 

\begin{figure}
\centering
\includegraphics[width=0.49\textwidth]{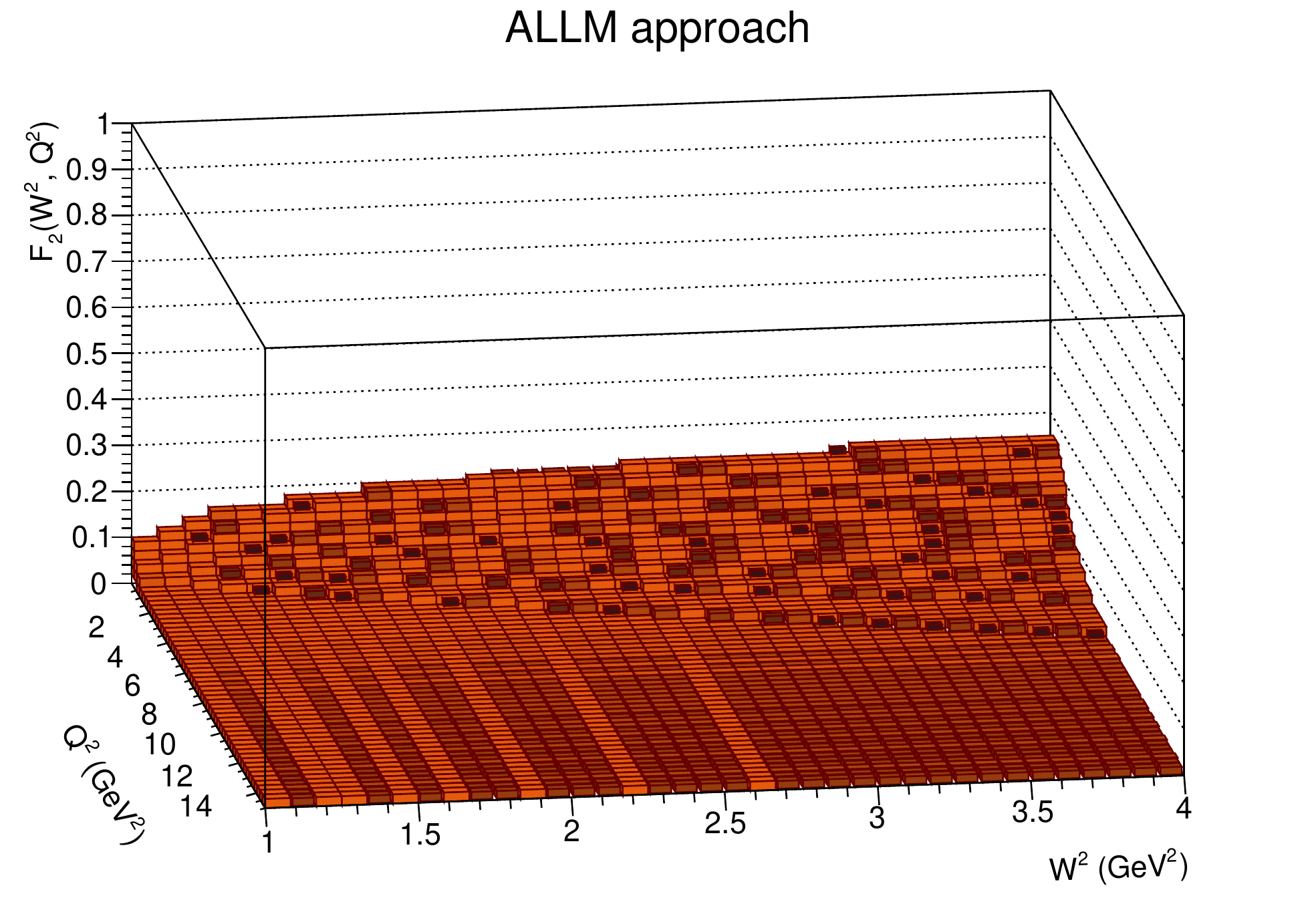}
\includegraphics[width=0.49\textwidth]{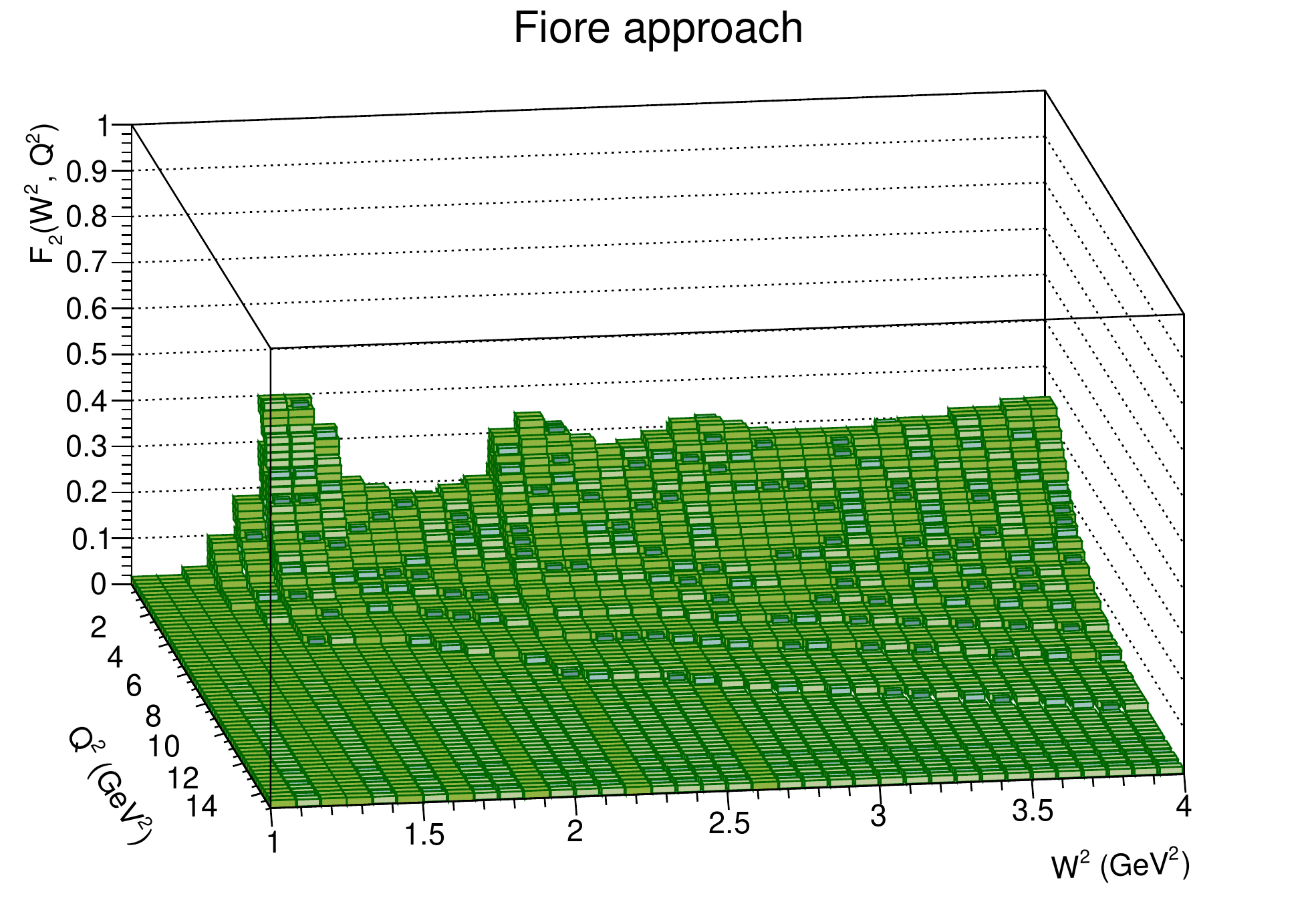}
\includegraphics[width=0.49\textwidth]{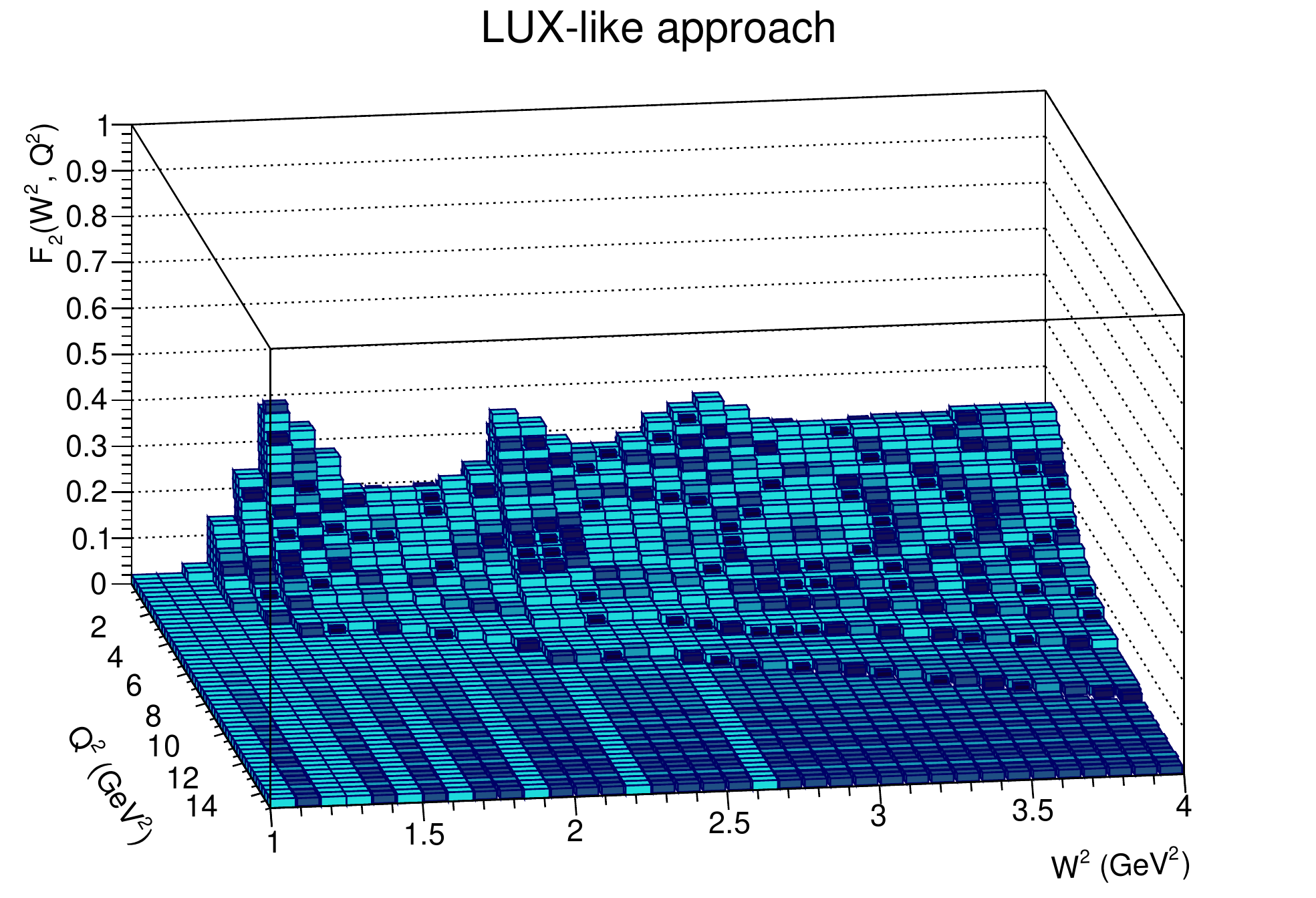}
\includegraphics[width=0.49\textwidth]{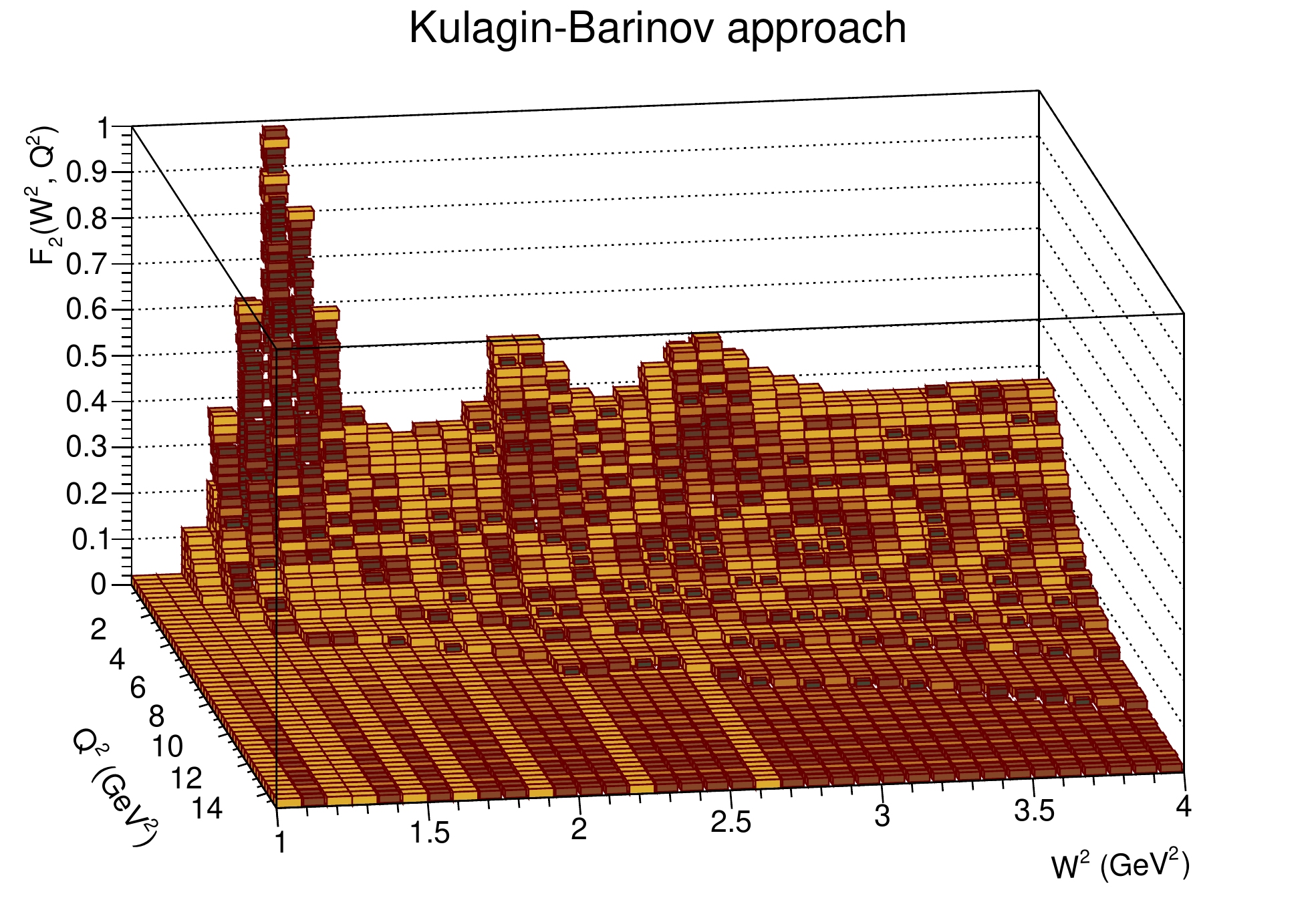}
\caption{\label{fig:F2_funs} Different parametrizations of structure functions depending on $W^{2}$ and $Q^{2}.$
}
\label{fig:F_2}
\end{figure}

In Fig.\ref{fig:F_2} we show the four different parametrizations of the structure function $F_2$ of the proton in the $(W^2,Q^2)$-plane. 
Here 
\begin{eqnarray}
W^2 = {1 - x_{\rm Bj} \over x_{\rm Bj}} Q^2 + m_p^2 \, 
\end{eqnarray}
is the $\gamma^* p$ cm-energy squared, so that $W$ is the invariant mass of the hadronic final state.

Here we observe, that the ALLM parametrization does not show the prominent resonance structures at low invariant mass. Indeed it is constructed in the spirit of parton-hadron duality and represents rather an ``averaged'' $F_2$. The remaining three parametrizations all contain explicit resonances, which are especially visible at low $Q^2$.

\section{Results}
In Table \ref{tab:widgets} we show integrated cross sections for different categories of $\gamma \gamma$ processes shown in Fig.\ref{fig:diagrams} for two different mass regions corresponding to the ALICE Collaboration results: low-mass region (LMR), where $0.5 < M_{ee} < 1.1$  GeV and intermediate-mass region (IMR) for $1.1 < M_{ee} < 2.7$ GeV.

\begin{table}
\centering
\begin{tabular}{l | r | r }
Structure function approaches  & $\sigma_{LMR}$ (nb)   & $\sigma_{IMR}$ (nb)  \\\hline
elastic & 2938.72 & 507.04\\
LUX-like & 346.53 & 191.40  \\
Kulagin - Barinov & 387.93  & 205.27 \\
Fiore & 653.07 & 347.08  \\
ALLM & 329.72 & 179.07 \\  
\end{tabular}
\caption{\label{tab:widgets}Total cross section for both mass region and different approaches.}
\end{table}

Cross sections as a function of some variables are presented separately for two different mass regions corresponding to the ALICE Collaboration regions: low-mass region (LMR) and intermediate-mass region (IMR).

The elastic contribution gives much larger contribution, especially for LMR. The ALLM, LUX-like and Kulagin-Barinov parametrization, although  differing in some regions of the $(x_{\rm Bj},Q^2)$ space give similar predictions for 
the integrated cross section.

\begin{figure}
\centering
\includegraphics[width=0.49\textwidth]{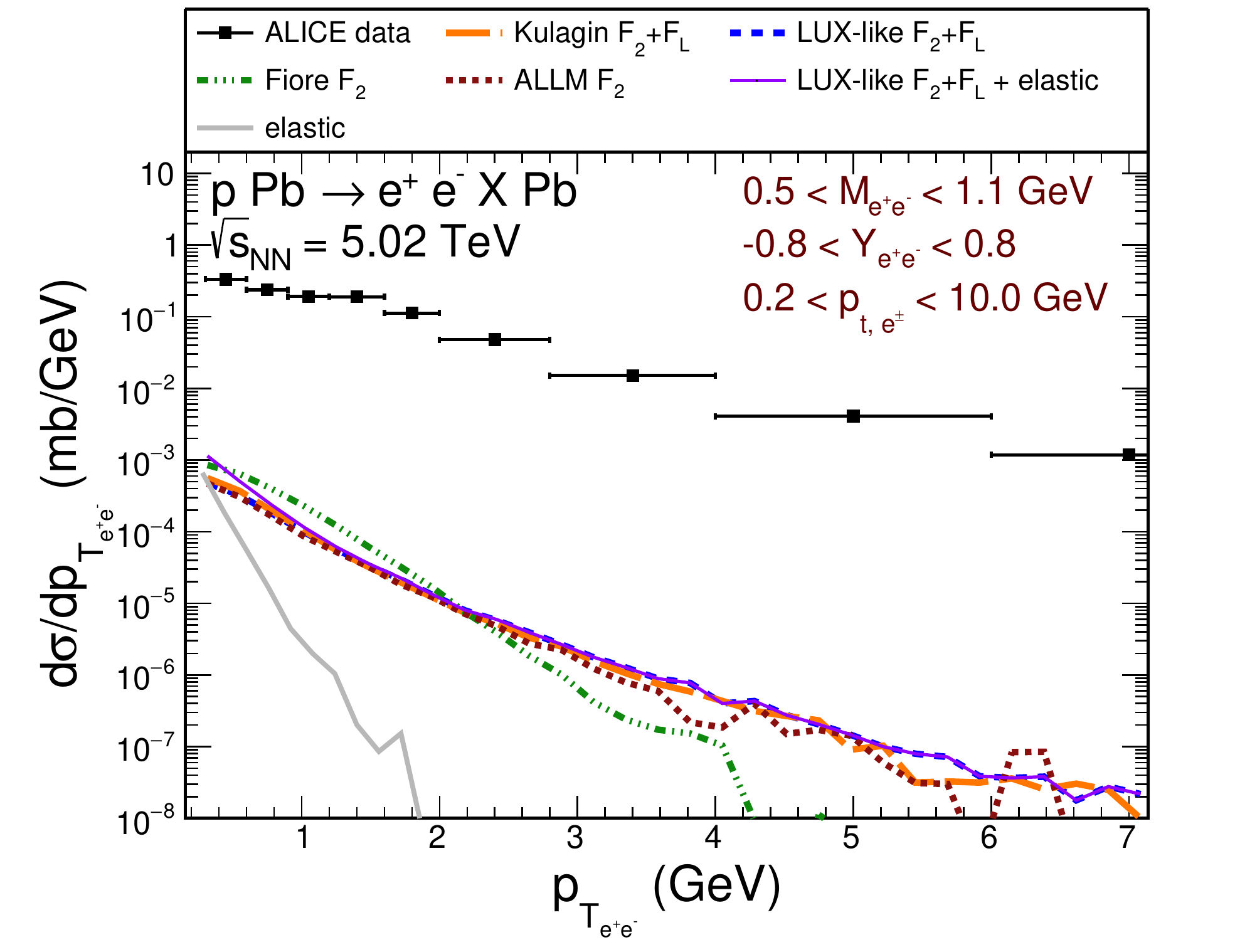}
\includegraphics[width=0.49\textwidth]{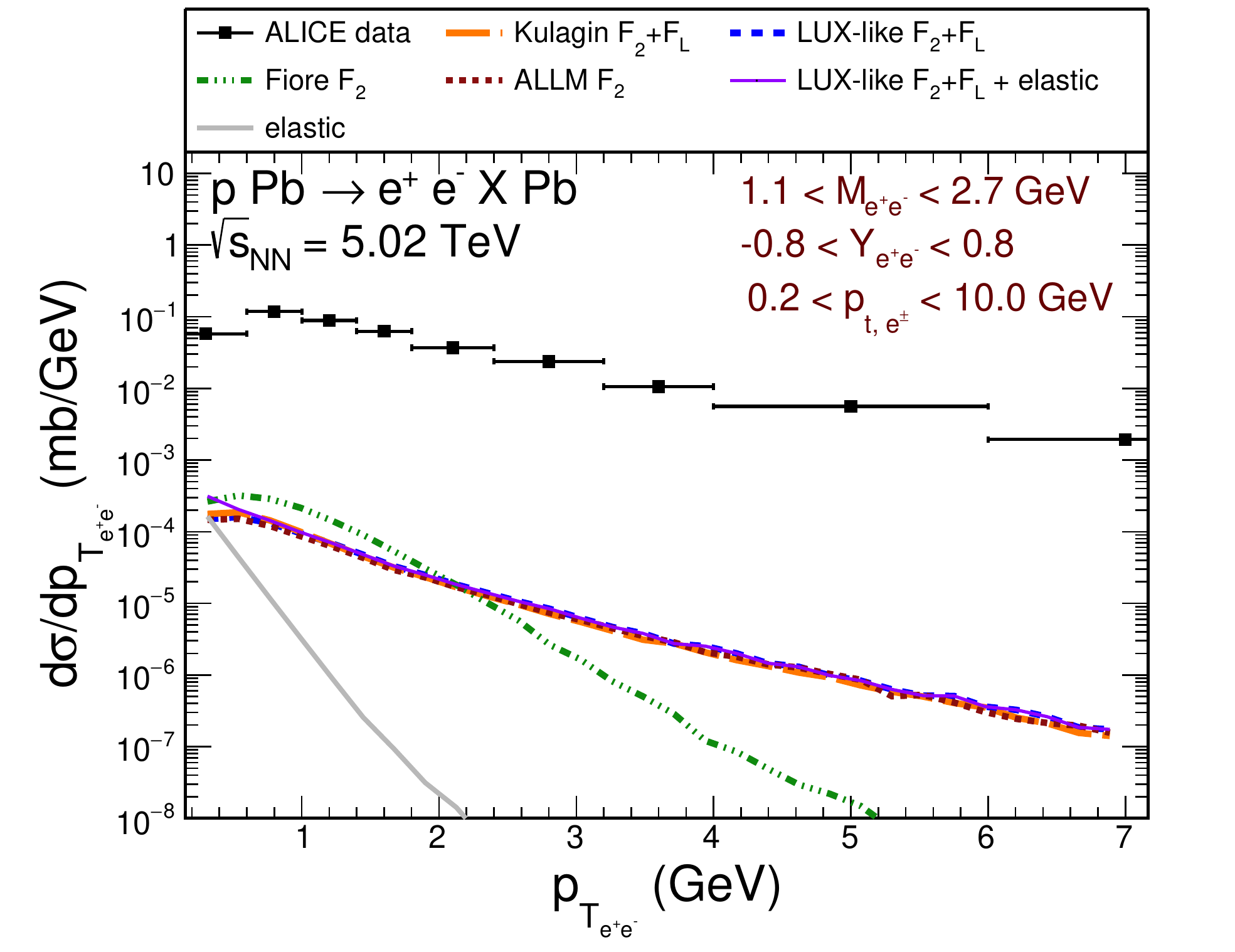}
\caption{\label{fig:ptpair}Distributions in $p_{T_{e^{+}e^{-}}}$ for LMR on the left and for the IMR on the right }
\end{figure}

Distributions in transverse momenta (see Fig.\ref{fig:ptpair}) correspond to the ALICE inclusive data. The $\gamma \gamma$ contribution is calculated here for the first time. It is much smaller than experimental ALICE data and contributions of other mechanism of dilepton production discussed e.g. in 
\cite{ALICE:2020mfy}.

Imposing an extra condition on rapidity gap one can select the $\gamma \gamma$ mechanism. Now we will concentrate therefore only on the $\gamma \gamma$ fusion. One can observe that the elastic contribution dominates at low lepton pair transverse momenta. The region of larger transverse momenta $p_{T,ee} >$ 1 GeV is dominated by the inelastic contribution. The differences for different parametrizations become visible for $p_{T,ee} >$ 3 GeV, where the cross section is rather small. It is not clear to us whether such a study will be possible within run 3 or run 4 of the LHC.

\begin{figure}
\centering
\includegraphics[width=0.49\textwidth]{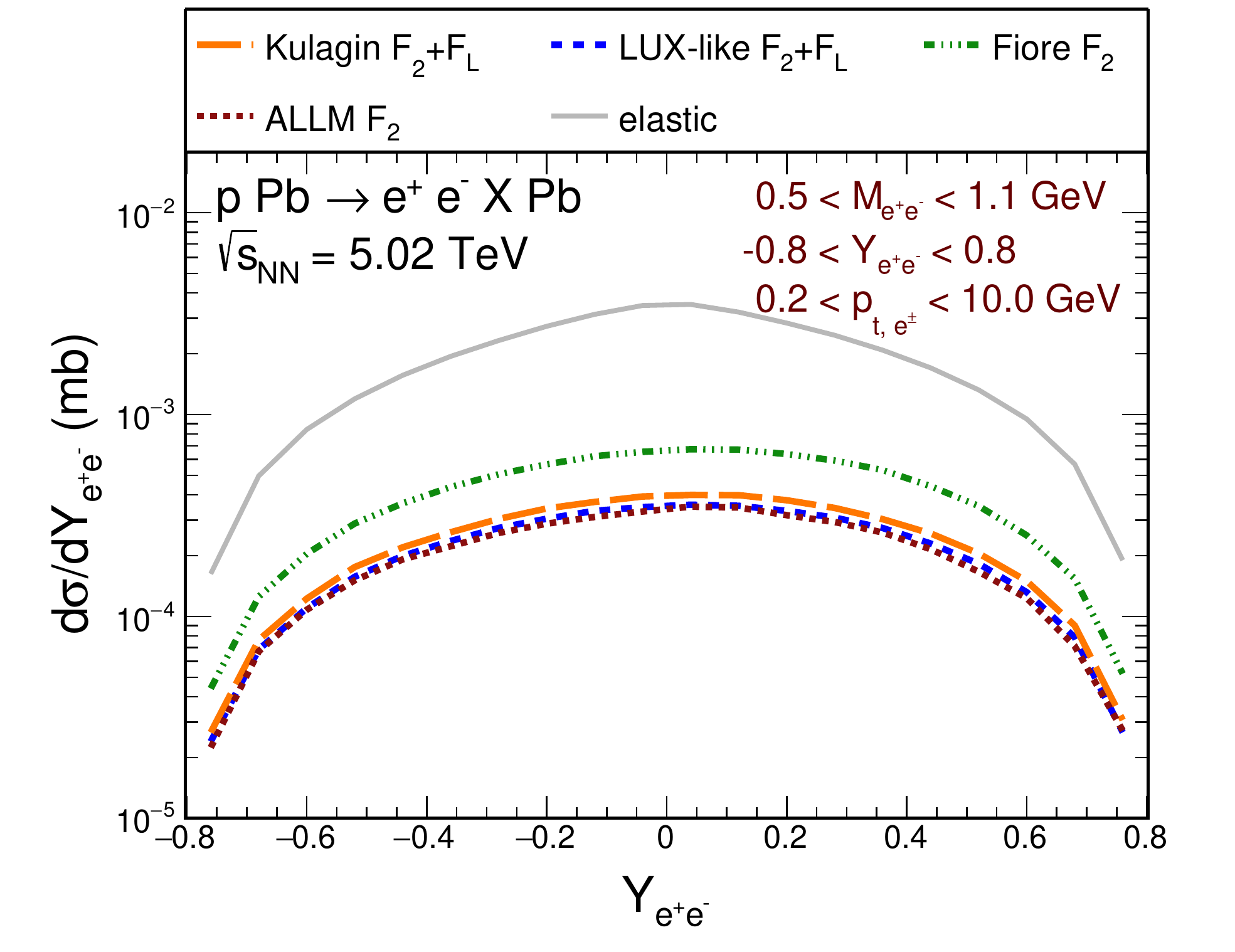}
\includegraphics[width=0.49\textwidth]{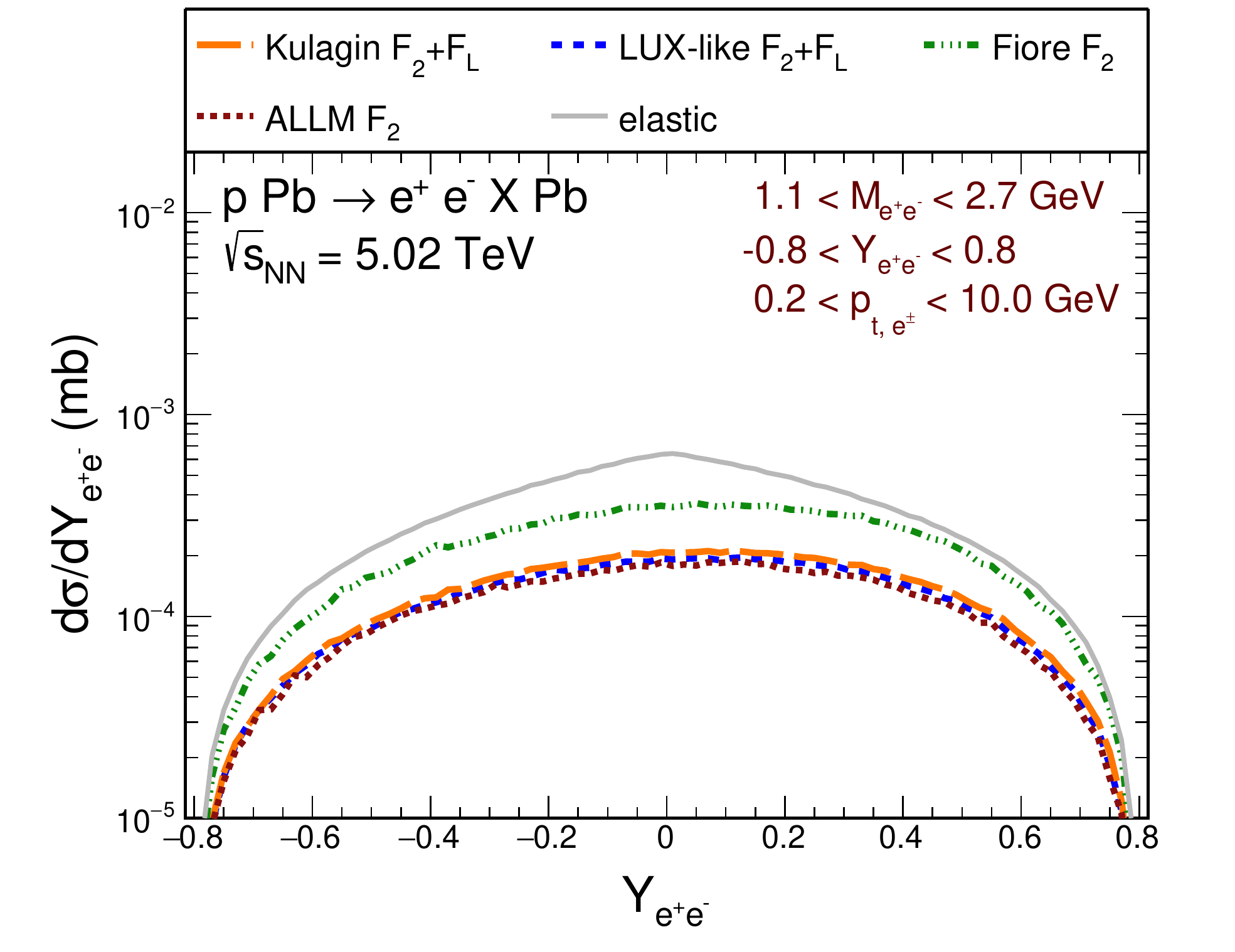}
\caption{\label{fig:Yee}Distributions in $Y_{e^{+}e^{-}}$ for LMR on the left and for the IMR on the right}
\end{figure}

The distributions in pair rapidity (within ALICE acceptance) are shown  in Fig.\ref{fig:Yee}. Here the three structure function parametrizations give very similar distributions.

\begin{figure}
\centering
\includegraphics[width=0.49\textwidth]{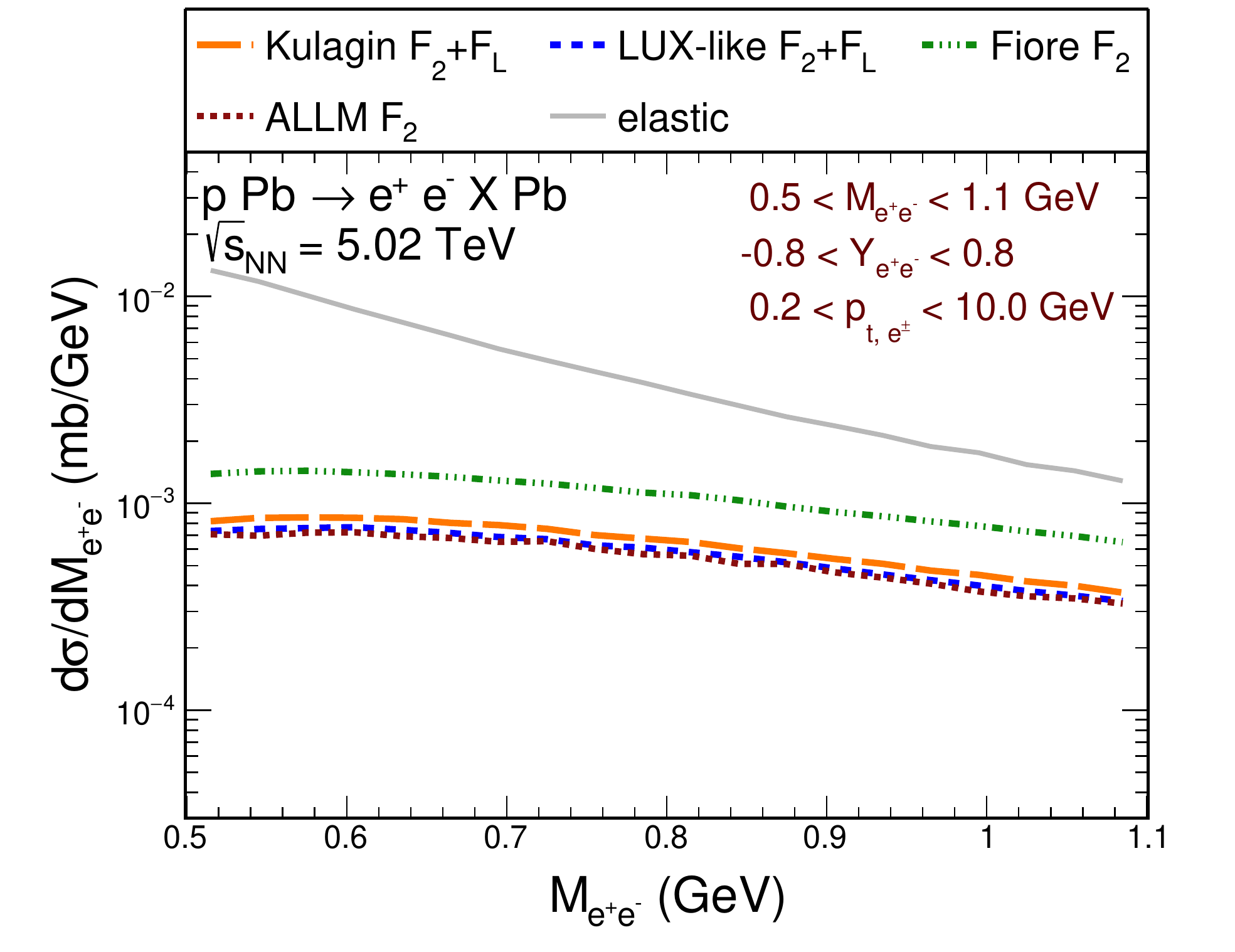}
\includegraphics[width=0.49\textwidth]{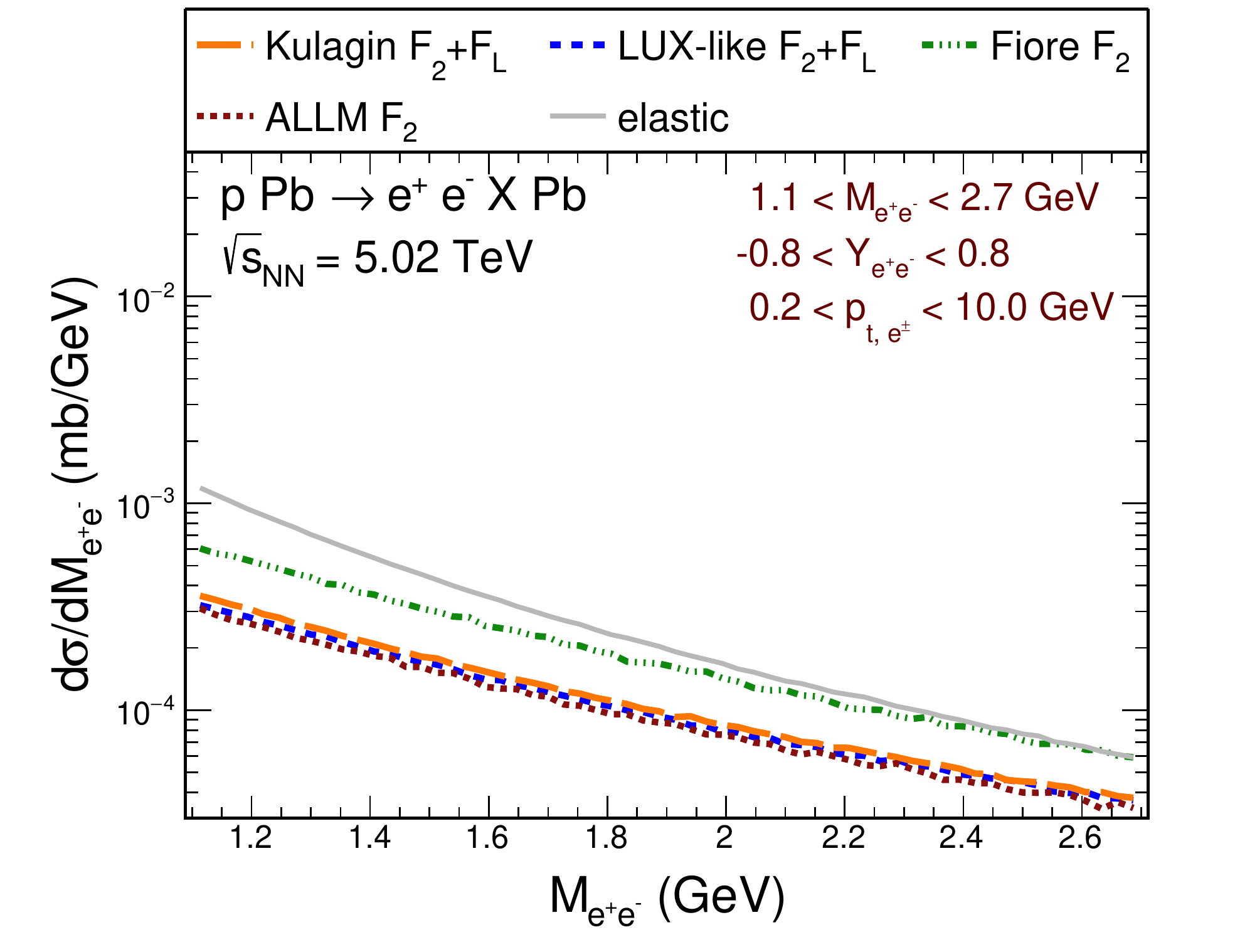}
\caption{\label{fig:Mll}Distributions in $M_{e^{+}e^{-}}$ for LMR on the left and for the IMR on the right.}
\end{figure}

In Fig.\ref{fig:Mll} we show the dielectron invariant mass distribution for the elastic-elastic and inelastic-elastic contributions. The first one gives a larger contribution than the second one. All structure functions, except of Fiore et al. \cite{Fiore2002ExplicitMR} give very similar distributions which gives confidence in our calculation.

\begin{figure}
\centering
\includegraphics[width=0.49\textwidth]{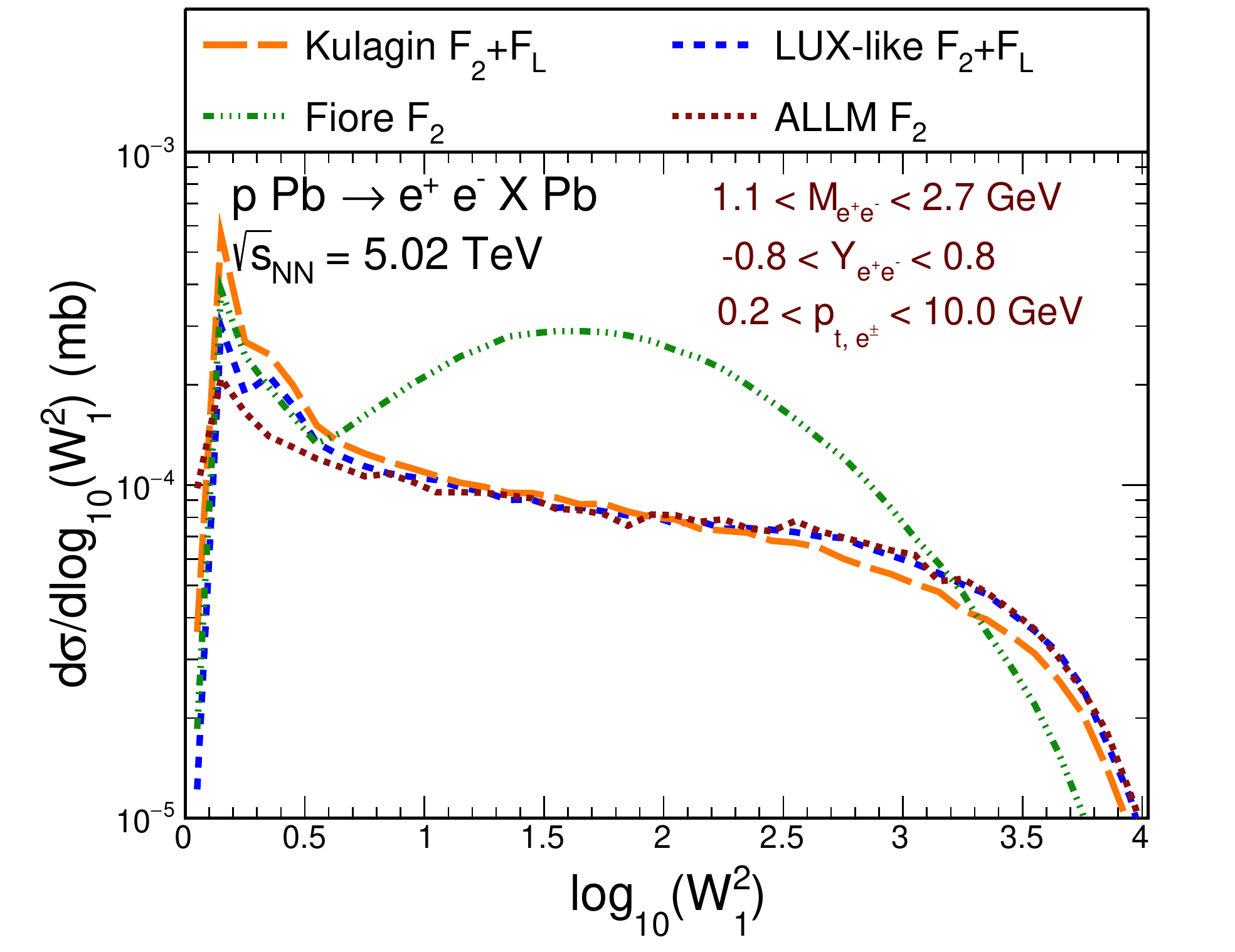}
\includegraphics[width=0.49\textwidth]{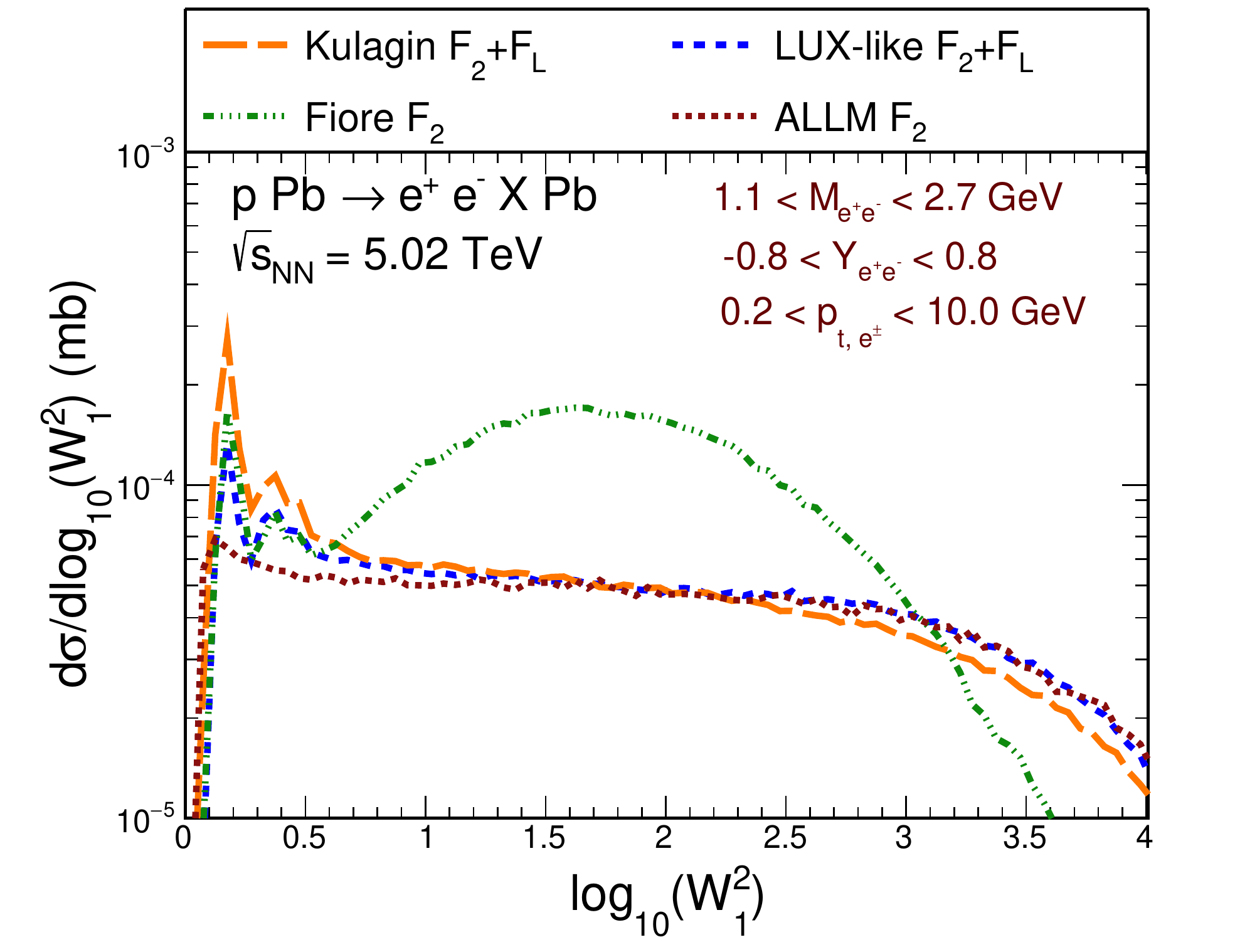}
\caption{\label{fig:logW2}Distribution in $log(W^{2})$ for LMR on the left and for the IMR on the right.}
\end{figure}

It is also interesting to inspect the rather theoretical distribution in photon-proton energy $W_1$. In Fig.\ref{fig:logW2} we show distributions rather in $\log_{10}W_1^2$, where $W_1^2 = \frac{Q_{1}^{2}}{x_{Bj_1}} -Q_{1}^{2}+m_p^2$, in order to cover the whole energy interval on one plot. Again the Fiore et al. parametrization \cite{Fiore2002ExplicitMR} gives quite different distribution. The results of other parametrization differ in the region of low $W_1$ where proton resonances occur. However, as discussed before, their contribution is not crucial for e.g. $Y_{e^+ e^-}$, $M_{e^+ e^-}$ distributions.

\begin{figure}
\centering
\includegraphics[width=0.49\textwidth]{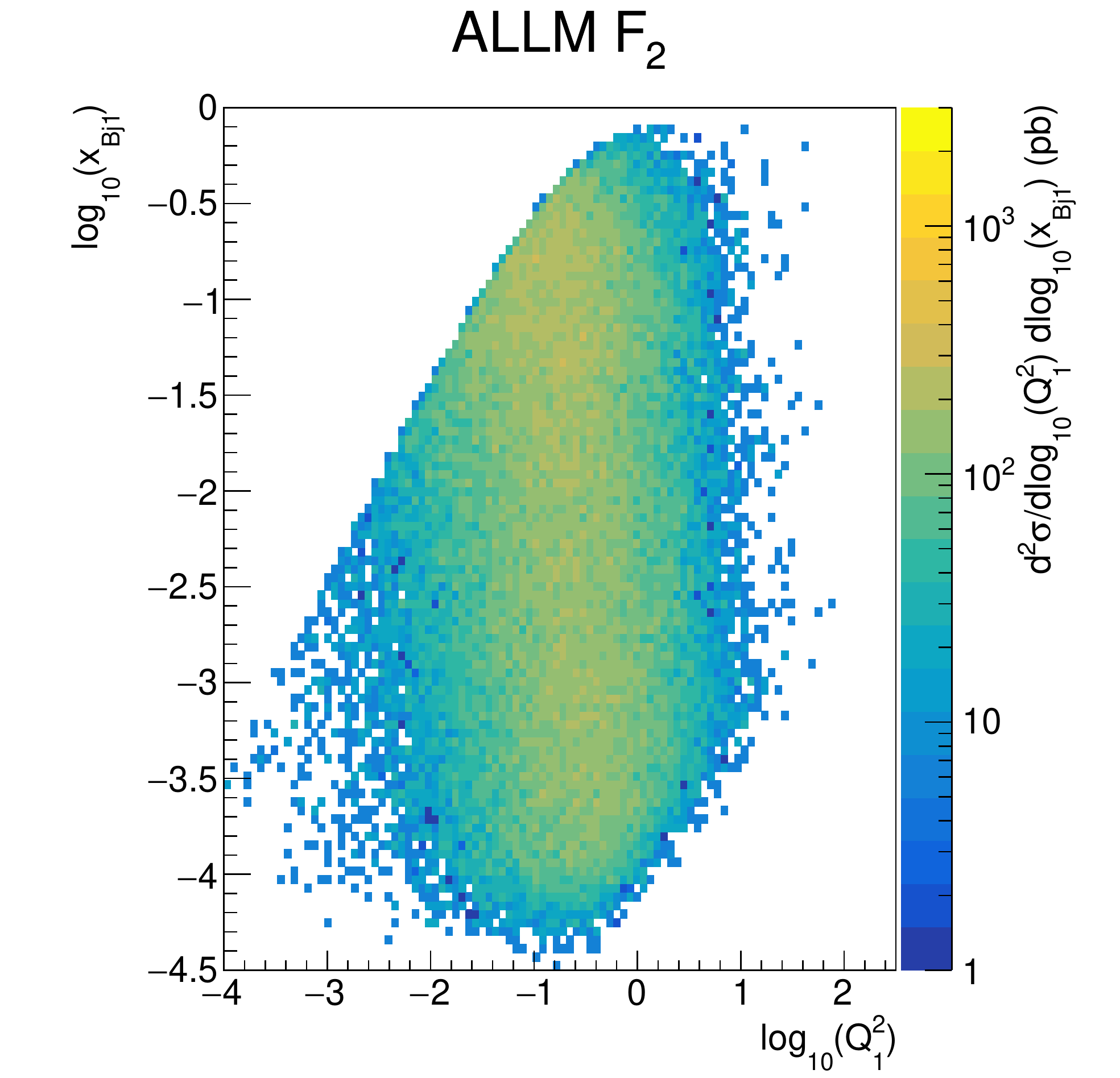}
\includegraphics[width=0.49\textwidth]{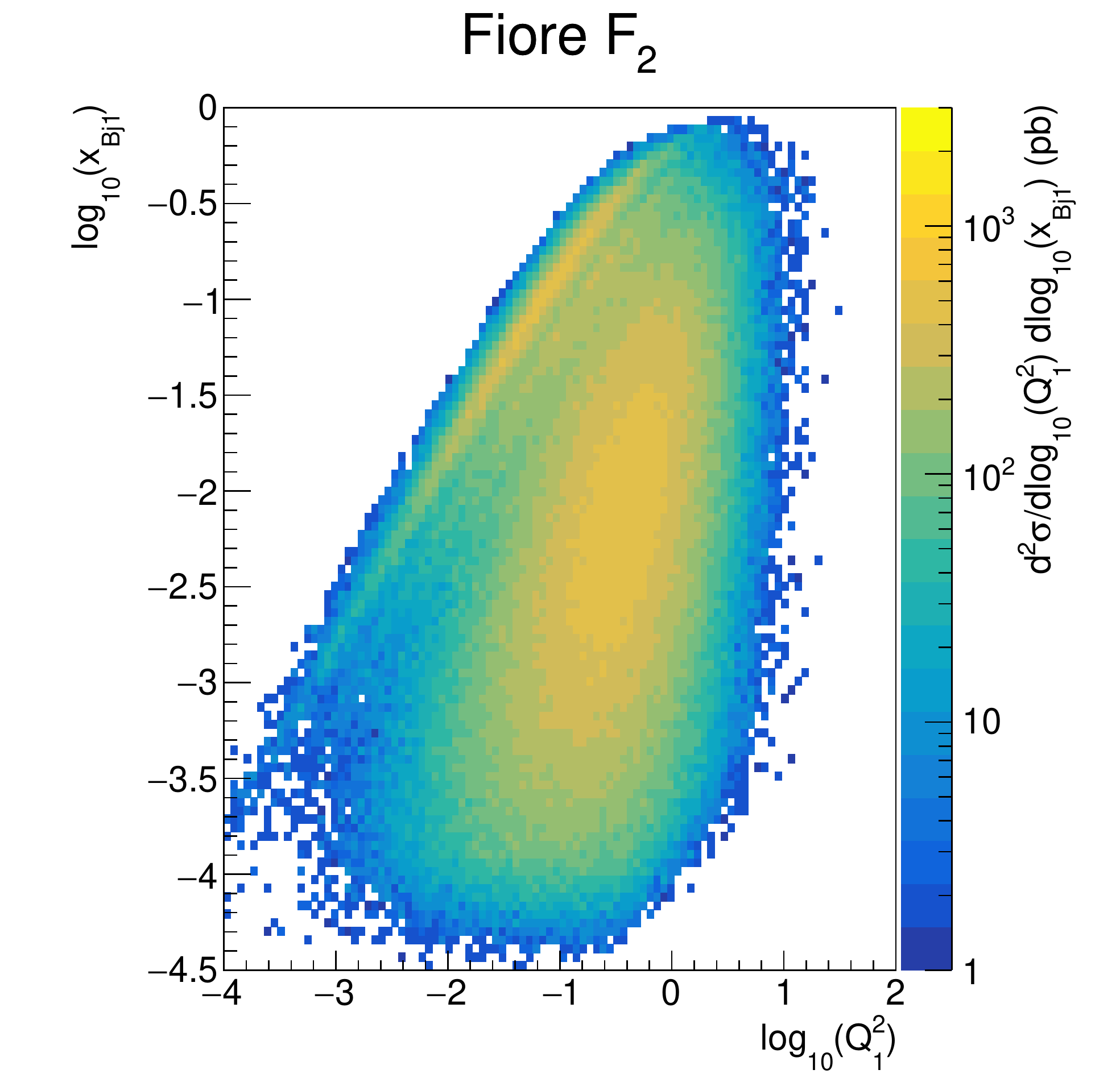}
\includegraphics[width=0.49\textwidth]{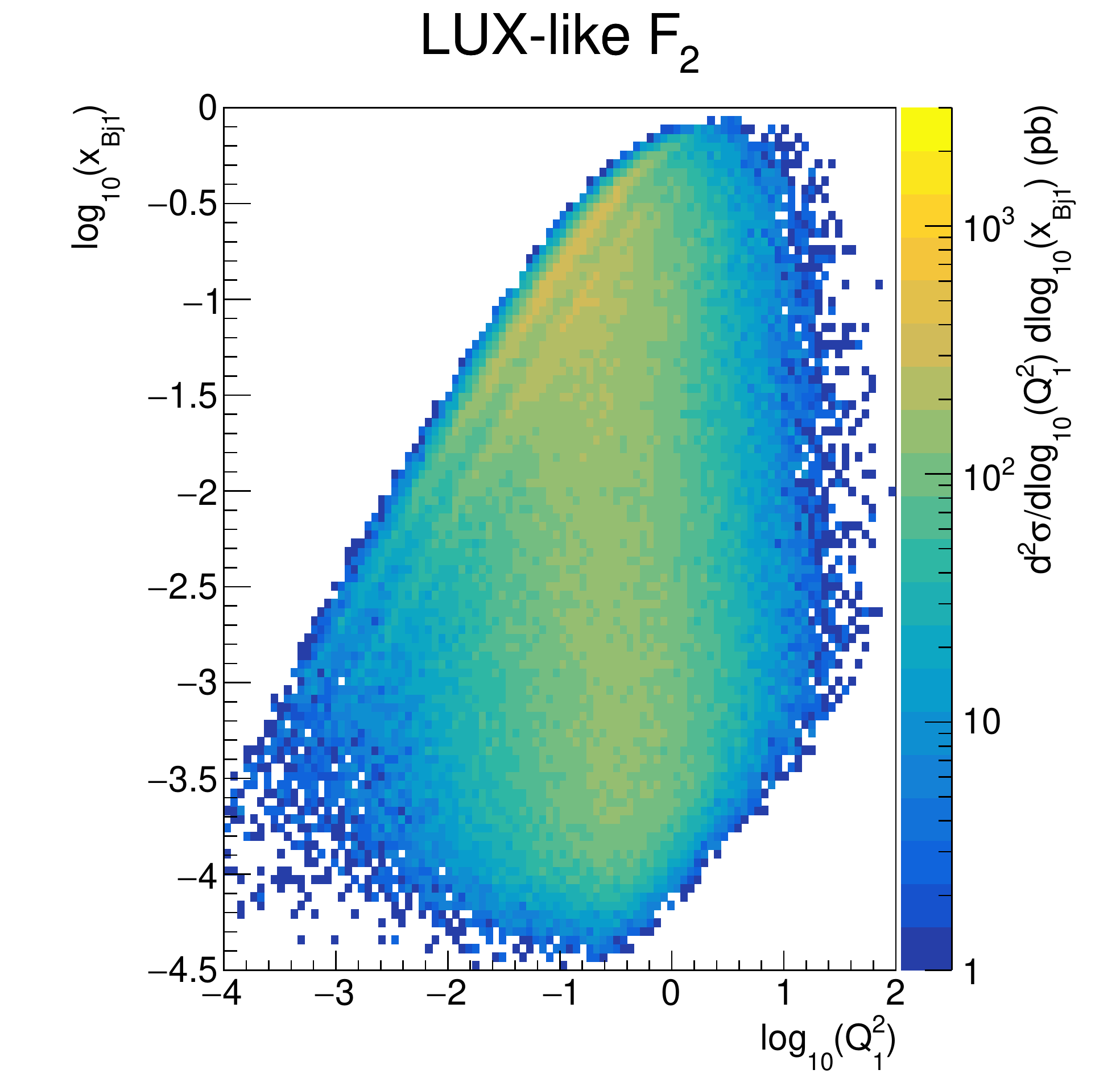}
\includegraphics[width=0.49\textwidth]{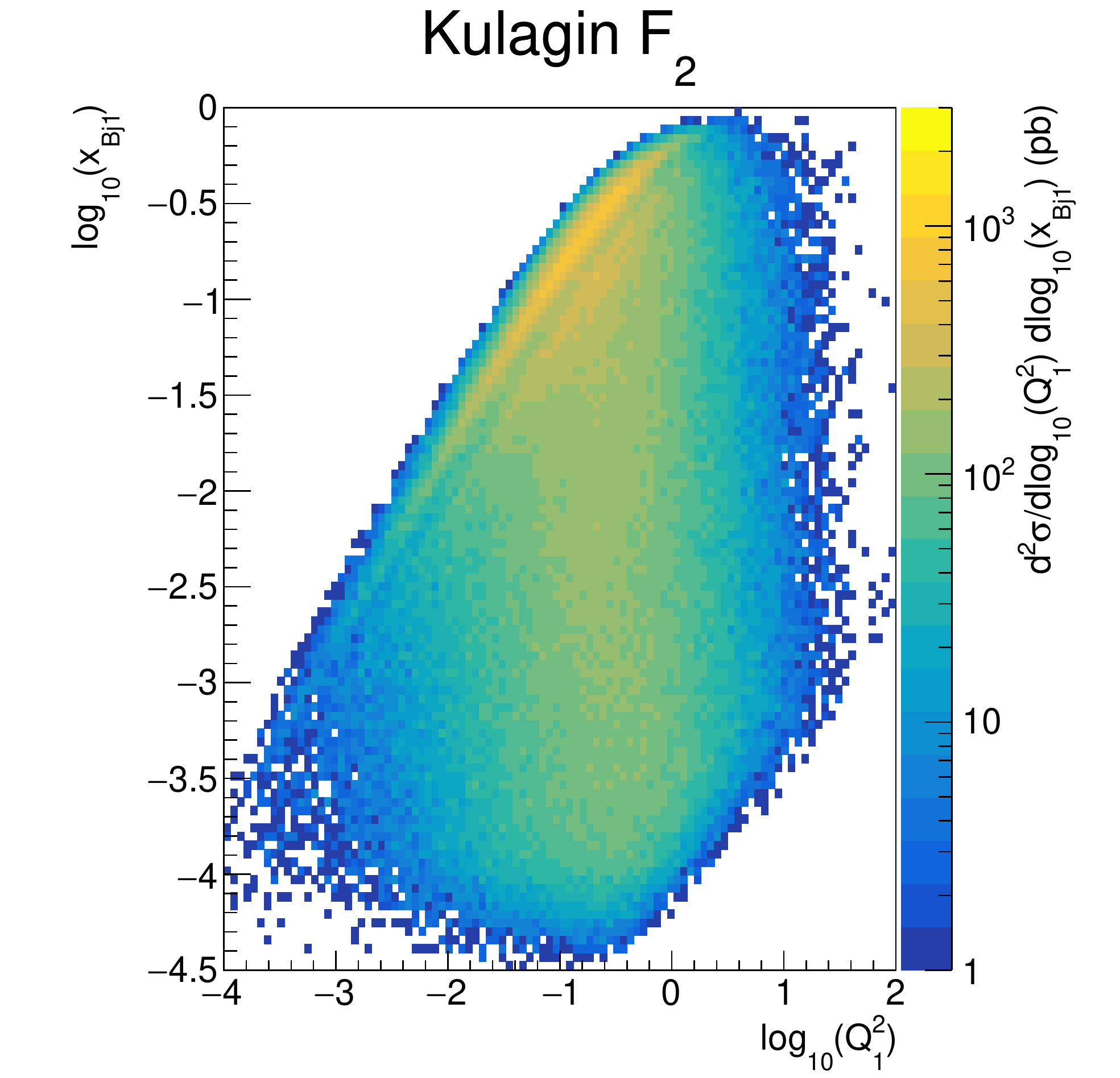}
\caption{\label{fig:LMR_logXbjQ2}Distribution in $\log_{10} x_{Bj}$ and $\log_{10} Q^{2}$ for four approaches of structure function: ALLM, Fiore, LUX-like and Kulagin respectively for LMR.}
\end{figure}

\begin{figure}
\centering
\includegraphics[width=0.49\textwidth]{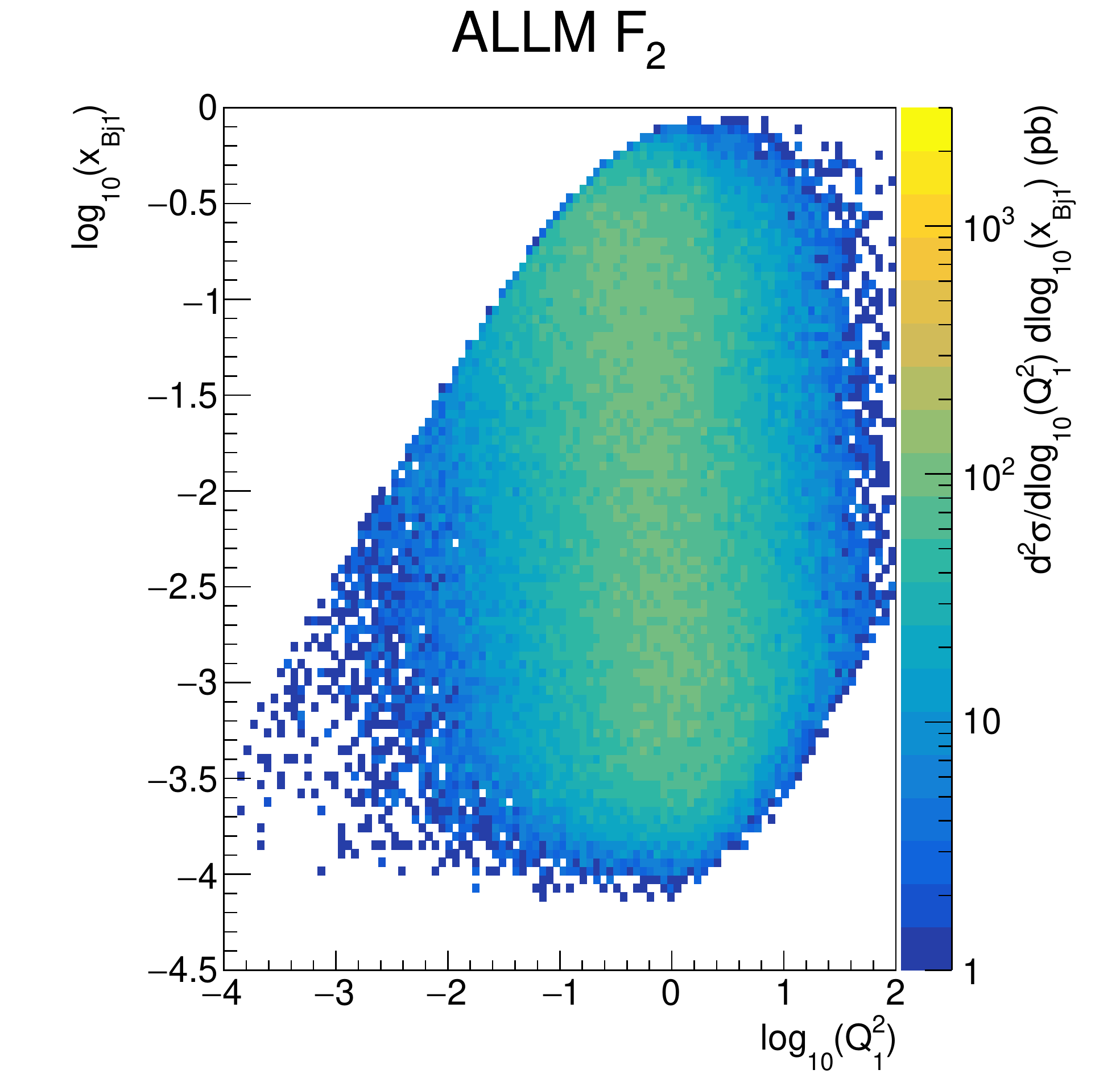}
\includegraphics[width=0.49\textwidth]{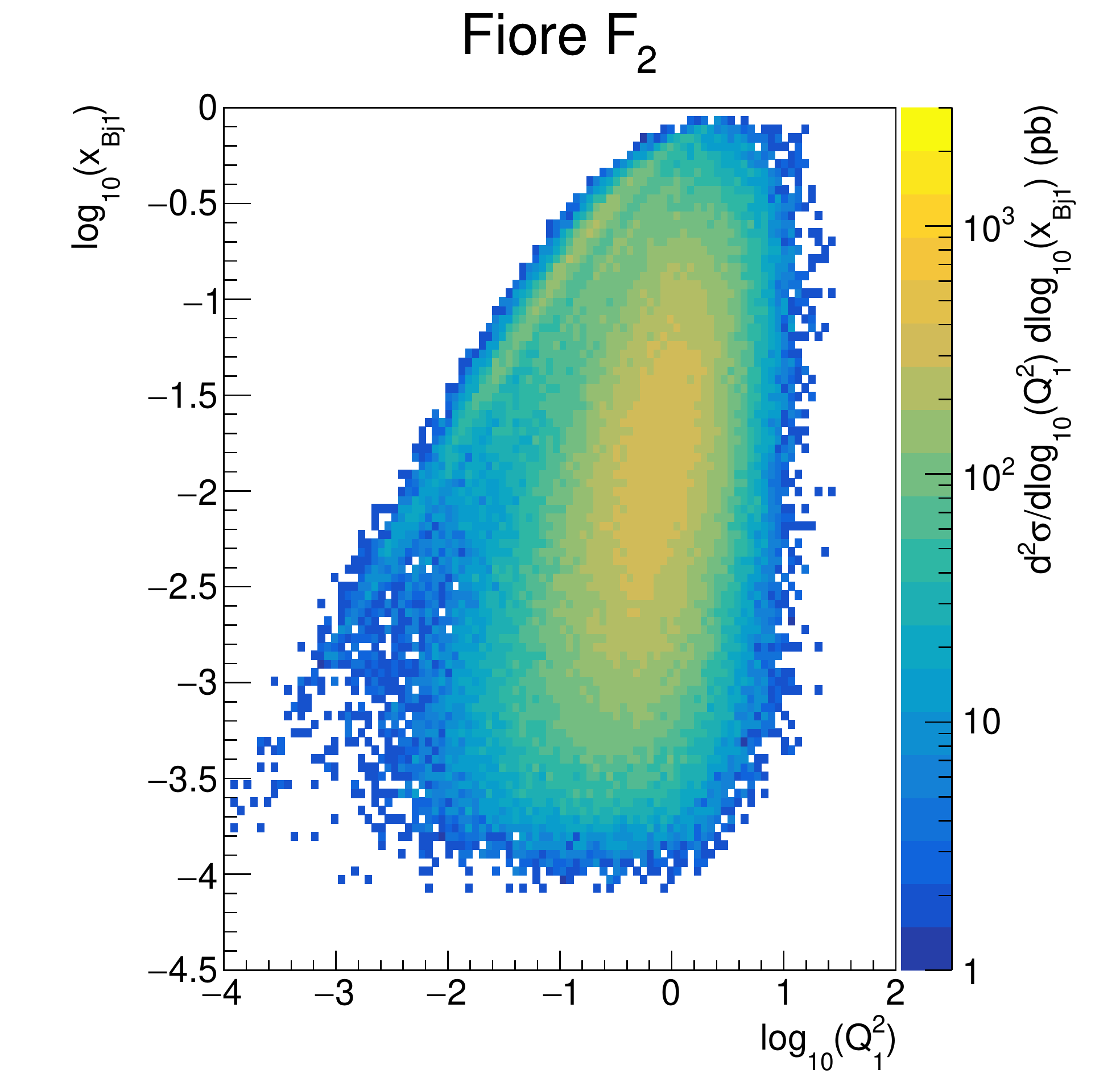}
\includegraphics[width=0.49\textwidth]{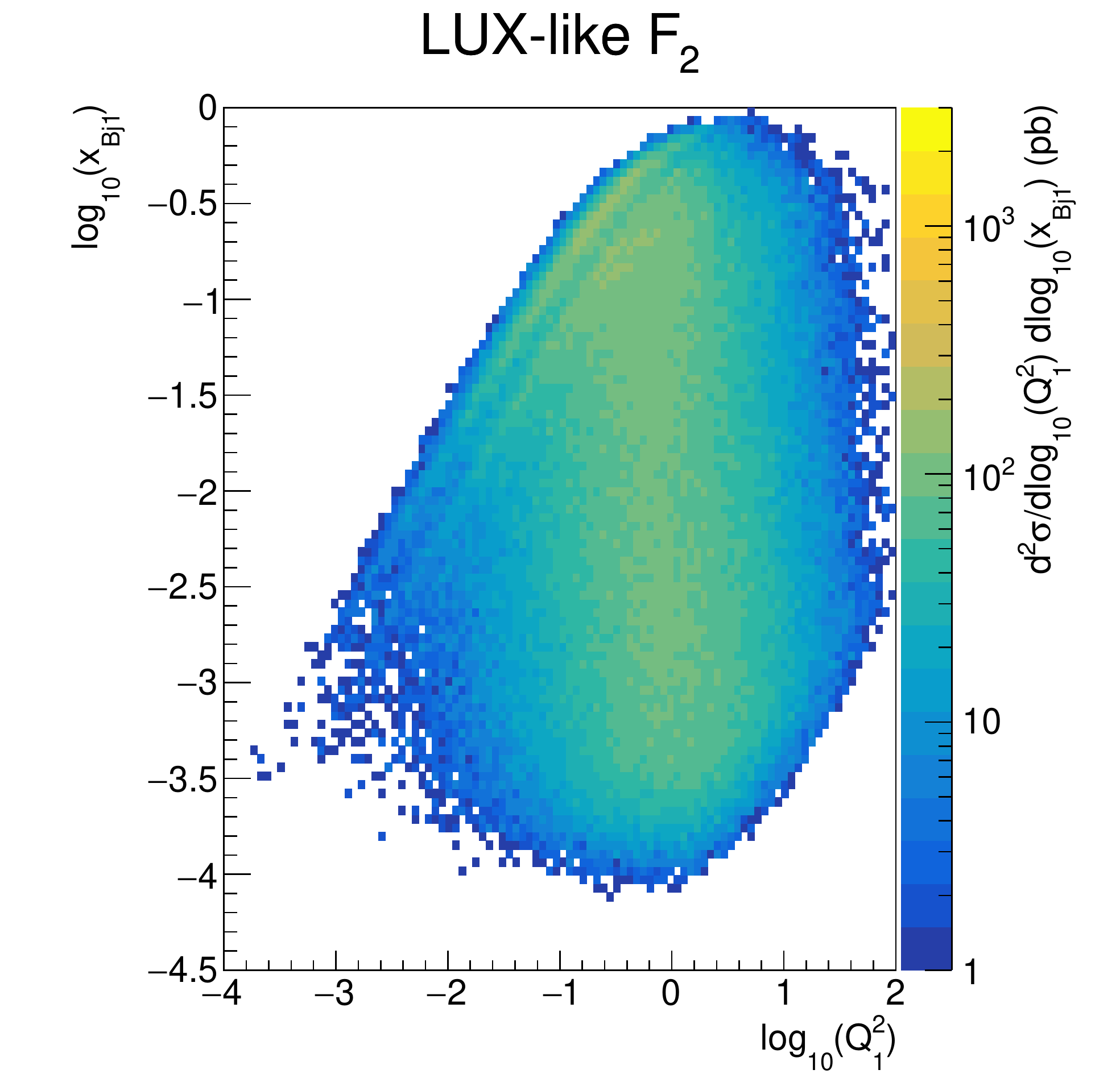}
\includegraphics[width=0.49\textwidth]{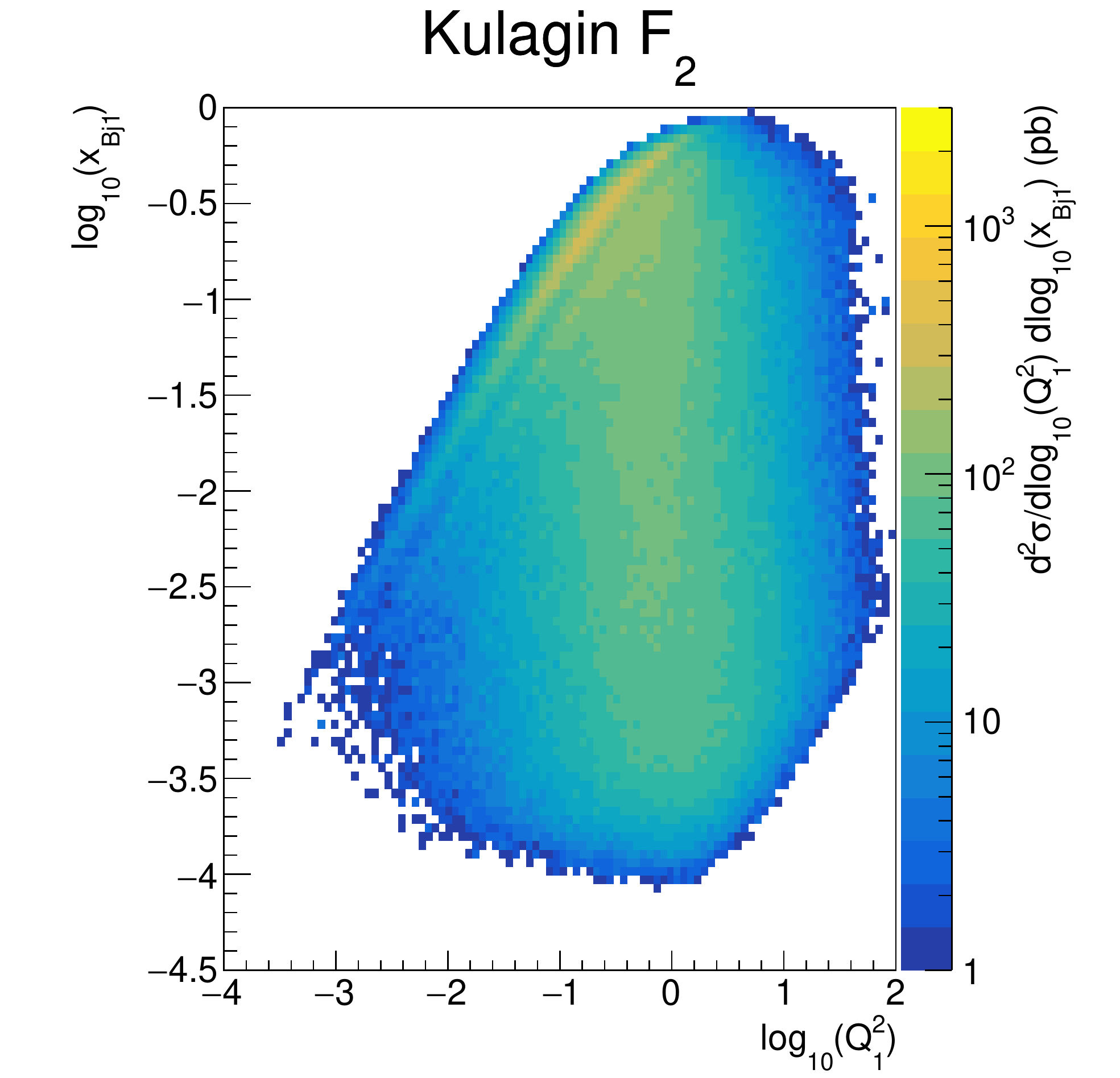}
\caption{\label{fig:IMR_logXbjQ2}Distribution in $\log x_{\rm Bj}$ and $\log Q^{2}$ for four approaches of structure function: ALLM, Fiore, LUX-like and Kulagin respectively for IMR.}
\end{figure}

Now we shall look more differentially. It is interesting to understand what regions of arguments of structure functions are important for the two-photon dilepton production. We start form the $(\log_{10} Q^2,\log_{10} x_{\rm Bj})$ distributions (see Fig.\ref{fig:LMR_logXbjQ2} and Fig.\ref{fig:IMR_logXbjQ2}). The figures show that our selected measurement with its specific cuts covers rather broad range of $x_{\rm Bj}$. A big part of the cross section comes from $Q_1^2 <$ 1 GeV$^2$, i.e. from clearly nonperturbative region, where partonic description breaks.  

\begin{figure}
\centering
\includegraphics[width=0.49\textwidth]{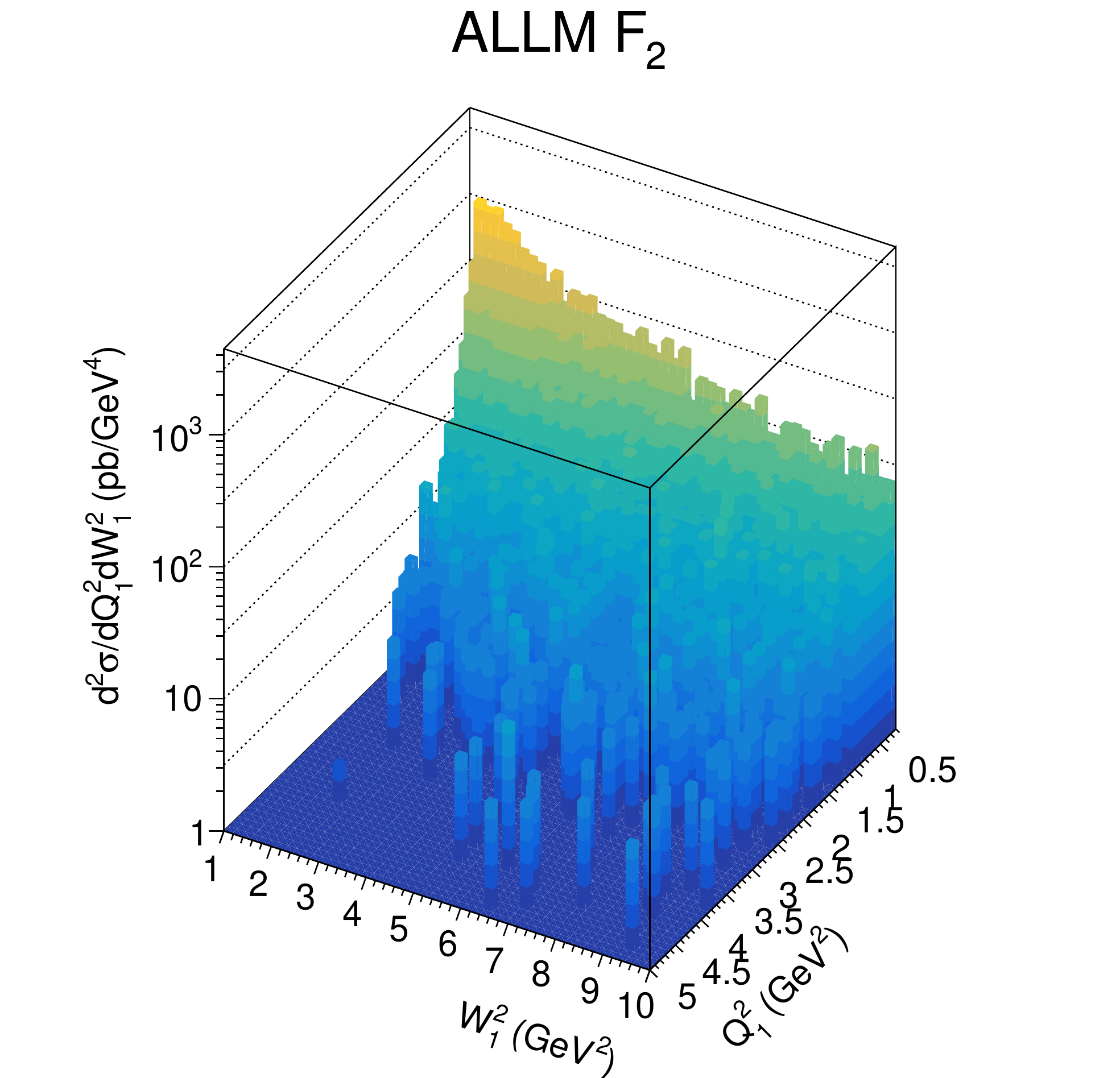}
\includegraphics[width=0.49\textwidth]{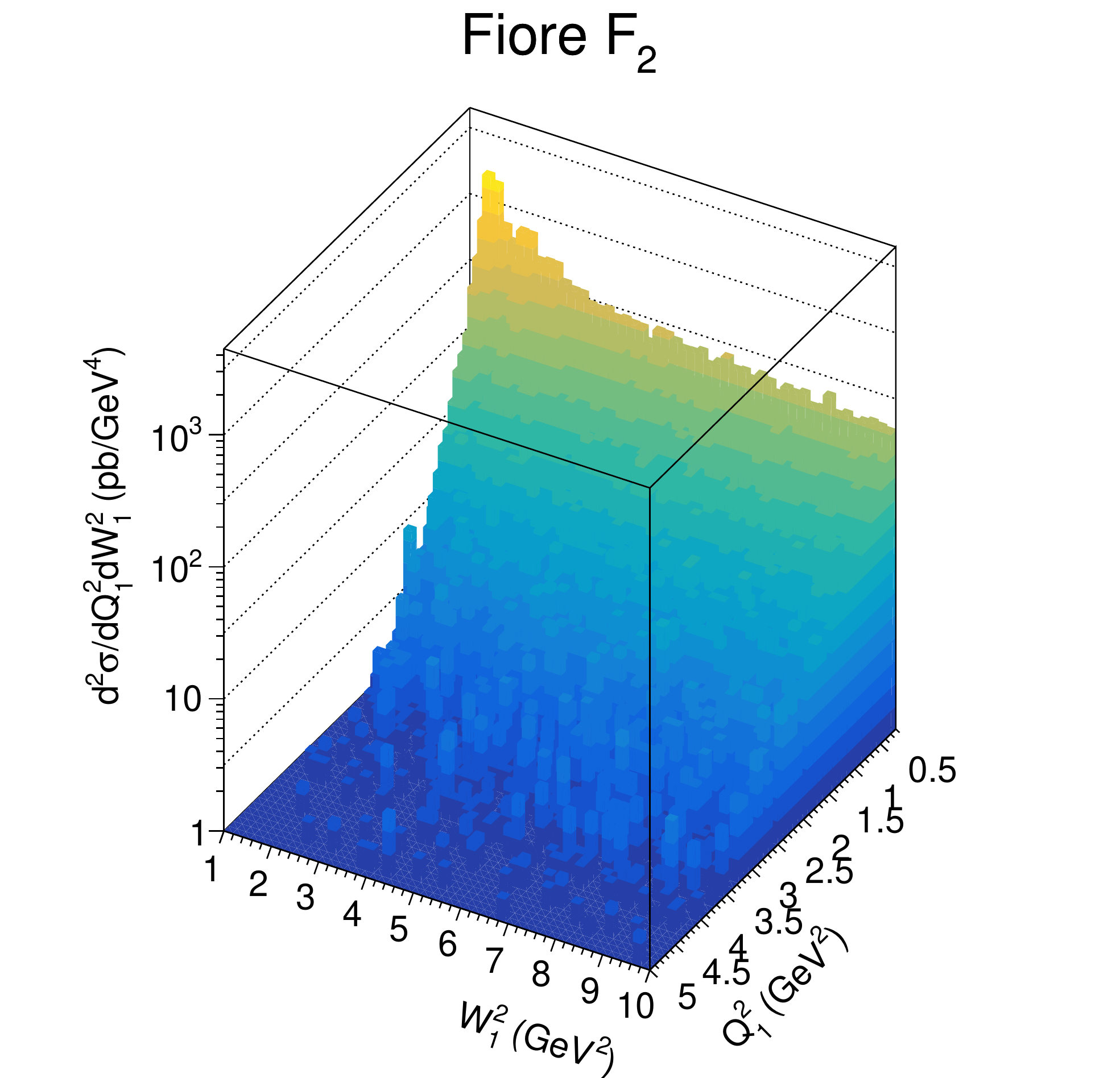}
\includegraphics[width=0.49\textwidth]{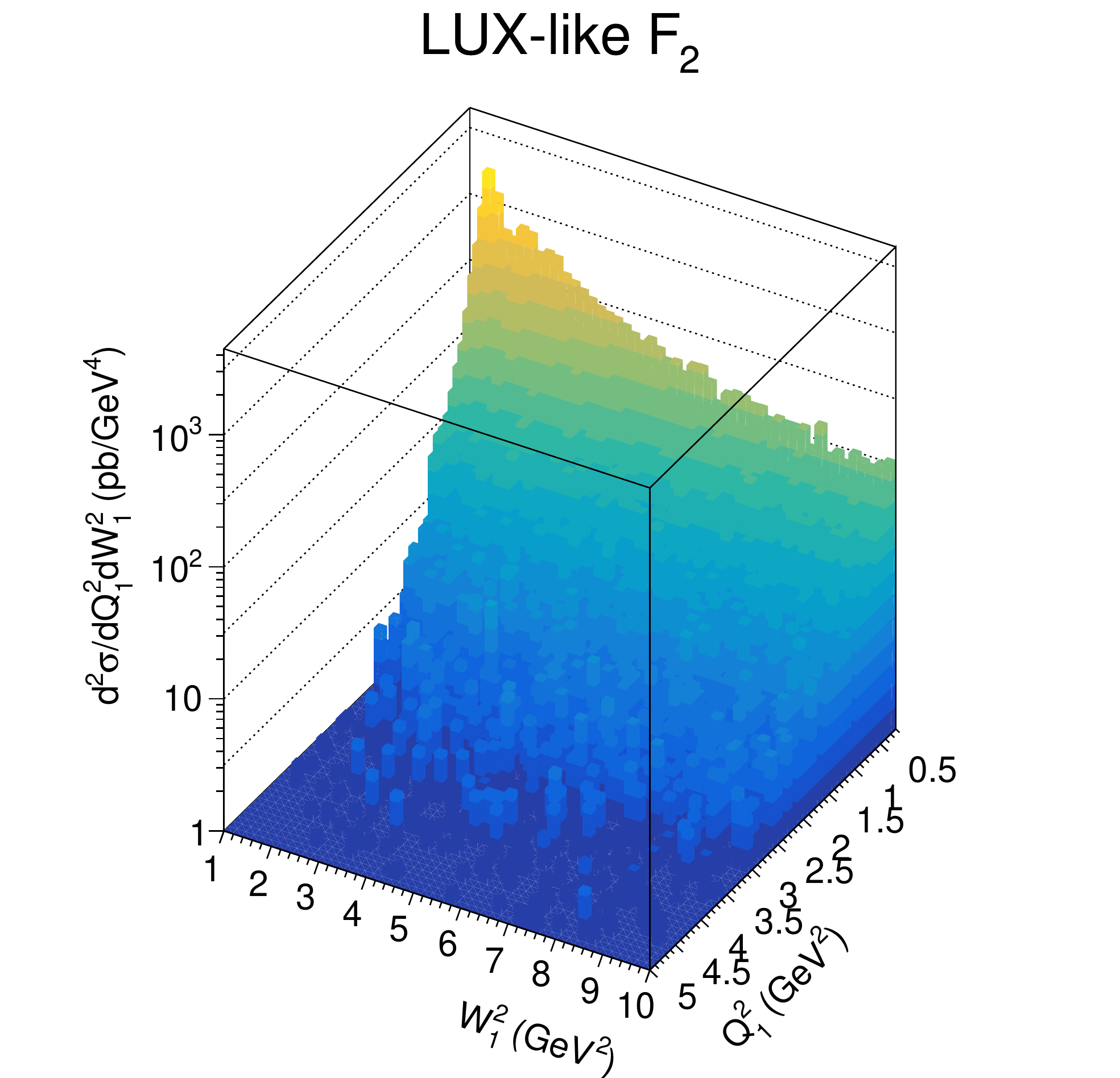}
\includegraphics[width=0.49\textwidth]{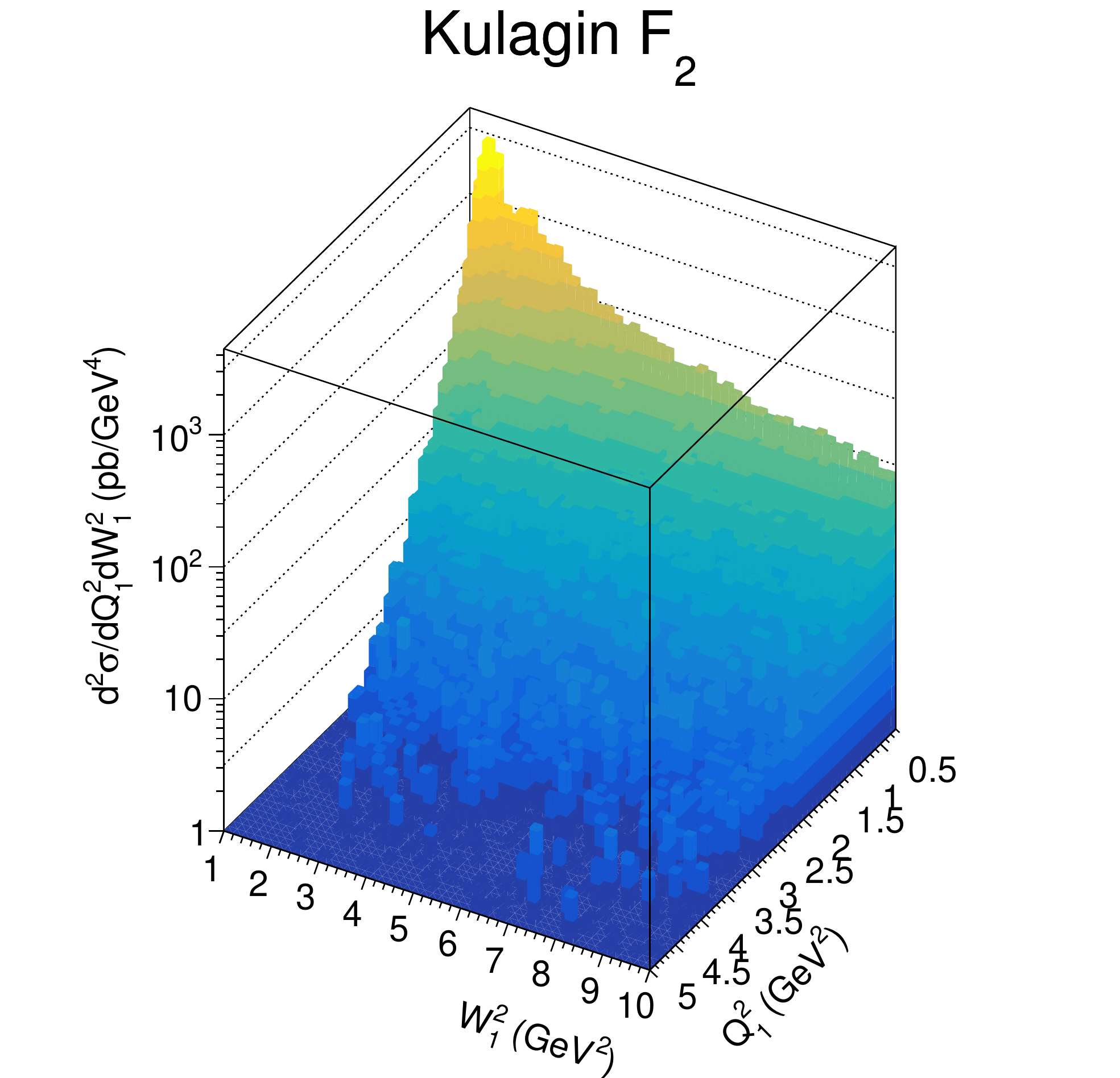}
\caption{\label{fig:LMR_W2_Q2}Distribution in $W^{2}$ and $Q^{2}$ for four approaches of structure function: ALLM, Fiore, LUX-like and Kulagin respectively for LMR.}
\end{figure}

\begin{figure}
\centering
\includegraphics[width=0.49\textwidth]{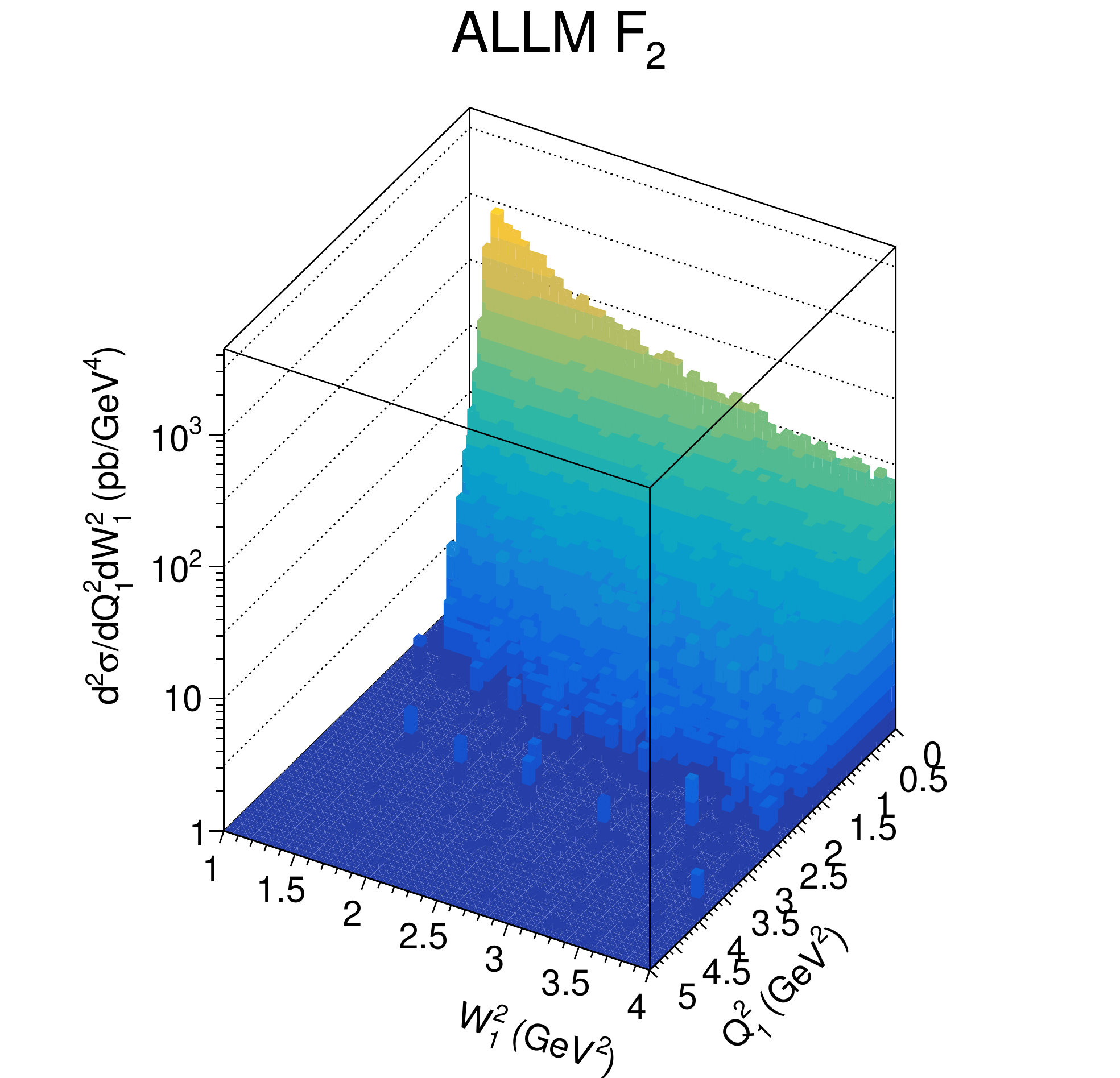}
\includegraphics[width=0.49\textwidth]{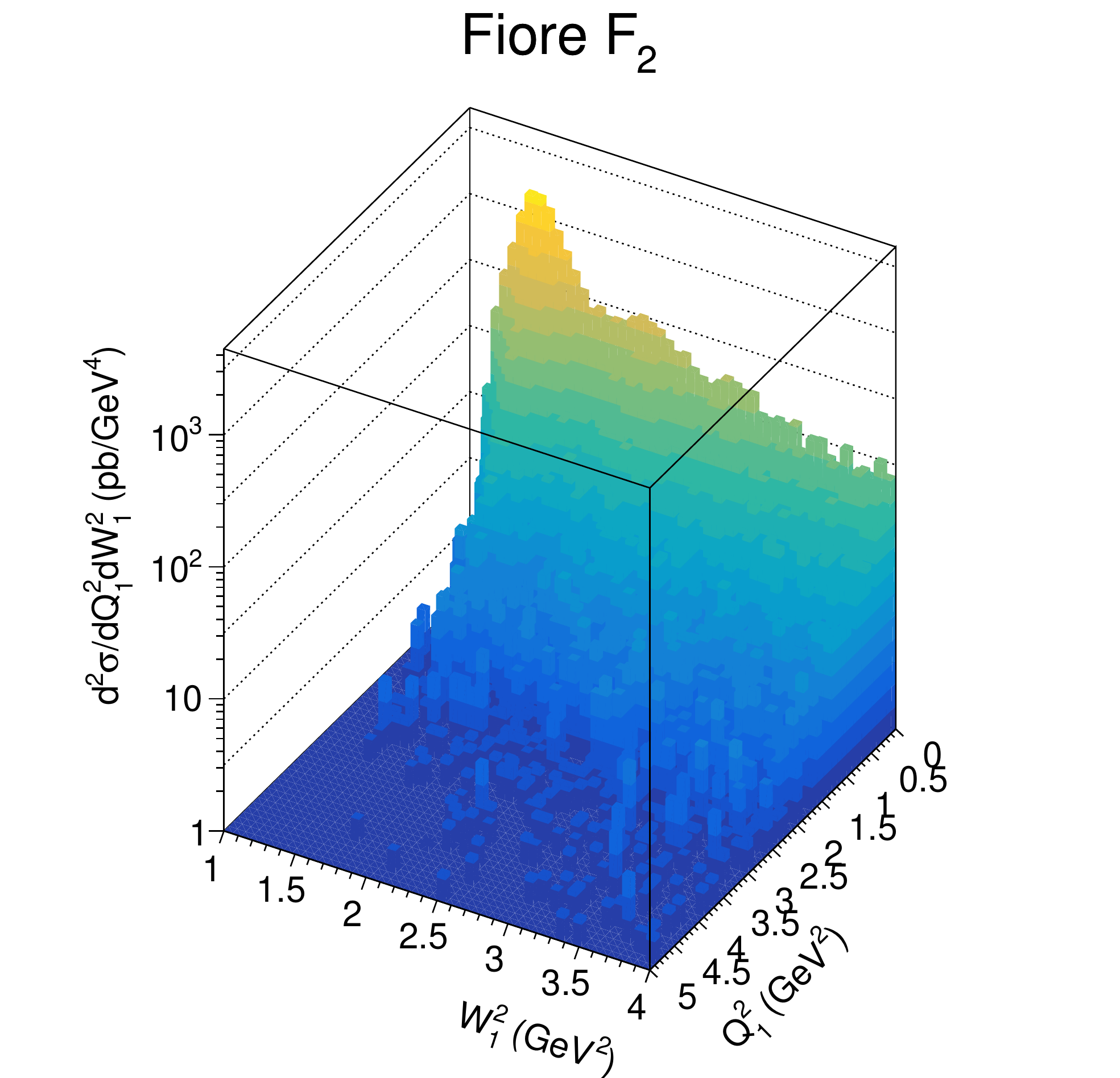}
\includegraphics[width=0.49\textwidth]{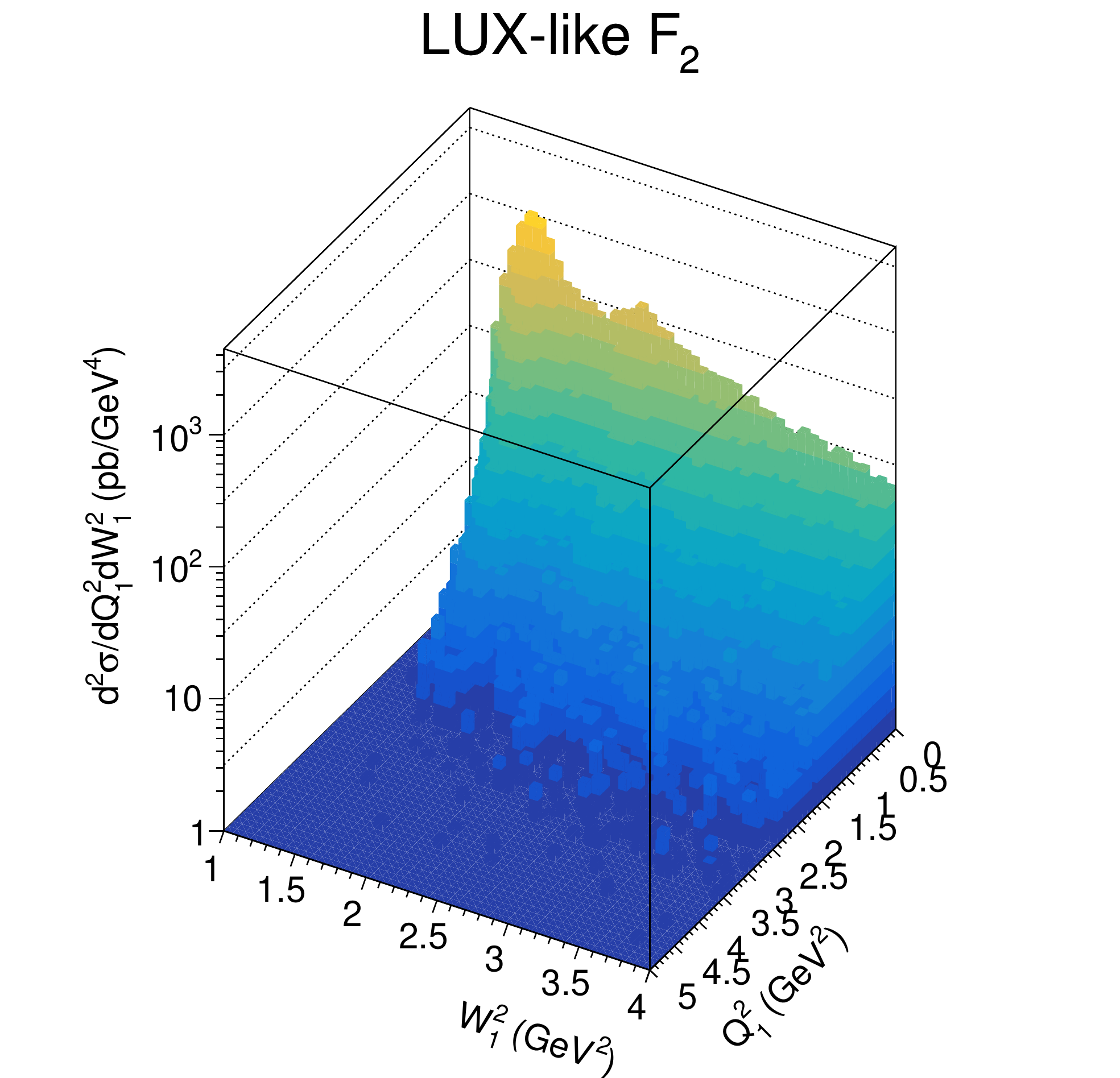}
\includegraphics[width=0.49\textwidth]{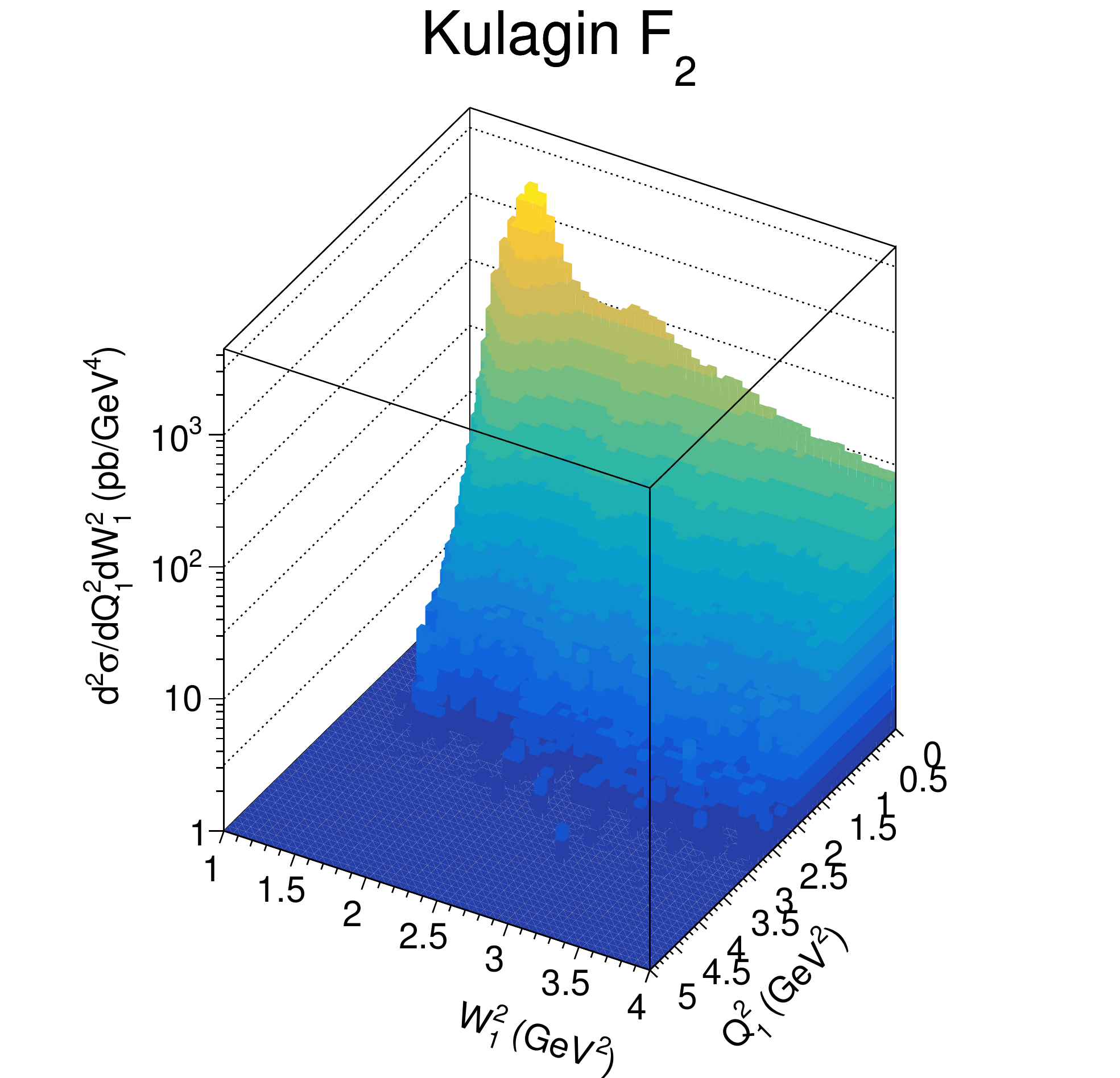}
\caption{\label{fig:IMR_W2_Q2}Distribution in $\log_{10} W^{2}$ and $\log_{10} Q^{2}$ for four approaches of structure function: ALLM, Fiore, LUX-like and Kulagin respectively for IMR.}
\end{figure}

Finally for completeness in Fig.\ref{fig:LMR_W2_Q2} and in Fig.\ref{fig:IMR_W2_Q2} we present also $(\log_{10}W_1^2,\log_{10}Q^2)$ distributions limiting to small range of $W_1$. Except for the ALLM parametrization, we clearly see contributions of individual resonances. We observe rather slight differences for different parametrizations.


\section{Conclusions}
In the present paper we have calculated the photon-photon contribution to the inclusive $e^+ e^-$ pair production in proton-Pb collisions. We have included processes when proton survives (elastic case) and when proton dissociates (inelastic case). The calculations have been performed in the so-called $k_t$-factorization approach including transverse momenta of intermediate photons. Modern parametrizations of proton structure functions have been used in the calculation.

The results have been compared to the existing data (distributions in transverse momentum of the dielectron pair) measured by the ALICE collaboration for two different windows on dielectron invariant mass, LMR ($0.5 < M_{ee} < 1.1$ GeV) and IMR ($1.1 < M_{ee} < 2.7$ GeV). We have checked that such a contribution is more than two orders of magnitude smaller than the published ALICE data. We conclude that the two-photon mechanism gives negligible contribution to the inclusive cross section.

The two-photon processes are interesting by themselves and could be studied in the future. This can be done by imposing rapidity gap veto.

We have performed the distributions in transverse momentum of the dilepton pair for different modern parametrizations of proton structure functions. We have shown that the region of relatively low dielectron masses (LMR+IMR) is sensitive to the nonperturabative regions (low-$Q^2$), and broad range of Bjorken-$x$. We have presented two-dimensional distributions in these variables ($\log_{10} Q^2,\log_{10} x_{Bj} $) and also in ($Q^2,W^2$). The second set of two-dimensional distributions shows that the ALICE kinematics could test also region of nucleon resonances, and actually a sizeable contribution to the distributions come from this region. Different parametrization used (Fiore et al, ALLM, Lux-like and Kulagin et al.) treat somewhat differently this domain of the structure functions. The Fiore et al. parametrization gives quite different result than the other used parametrizations. However, the Fiore parametrization was obtained from a fit to a rather narrow range of $Q^2$ and $W$ relevant for JLAB kinematics only. Extending this fit outside the JLAB region may be not justified.  

\bibliography{main}

\end{document}